\documentclass{aa}
\usepackage{amsmath,graphics,amssymb}
\usepackage{times}

\allowdisplaybreaks

\newfont{\eufont}{eufm10}
\def\eu #1{\mbox{\eufont #1}}

\newcommand{\dmdt}{\dot{\text{\eu M}}}

\newcommand{\vr}{{v_r}}
\newcommand{\pr}[1]{{#1}_{\mathrm{p}}}
\newcommand{\io}[1]{{#1}_{\mathrm{i}}}
\newcommand{\el}[1]{{#1}_{\mathrm{e}}}
\newcommand{\zav}[1]{\left(#1\right)}
\newcommand{\hzav}[1]{\left[#1\right]}

\newcommand{\rad}{\mathrm{rad}}

\newcommand{\de}{\mathrm{d}}
\newcommand{\Teff}{\mbox{$T_\mathrm{eff}$}}
\newcommand{\er}{{\mathrm e}}
\newcommand{\je}{{}^{[1]}}
\newcommand{\dv}{{}^{[2]}}
\newcommand{\tr}{{}^{[3]}}
\newcommand{\ct}{{}^{[4]}}
\newcommand{\pe}{{}^{[5]}}
\newcommand{\se}{{}^{[6]}}
\newcommand{\sm}{{}^{[7]}}
\newcommand{\os}{{}^{[8]}}

\newcommand{\Kriz}{K\v{r}\'{\i}\v{z}}

\begin{document}

\title{NLTE models of line-driven stellar winds}
\subtitle{I. Method of calculation and first results for O stars}
\titlerunning{NLTE models of line-driven stellar winds I.}

\author{J.  Krti\v{c}ka\inst{1,2,3} 
 \and J. Kub\'at\inst{3}}
\authorrunning{J. Krti\v{c}ka and J. Kub\'at}

\offprints{J. Krti\v{c}ka,\\  \email{krticka@physics.muni.cz}}

\institute{\'Ustav teoretick\'e fyziky a astrofyziky P\v{r}F MU,
            CZ-611 37 Brno, Czech Republic
           \and
           Department of Physics and Astronomy, University of Glasgow, 
           Glasgow G12 8QQ, UK
           \and
           Astronomick\'y \'ustav, Akademie v\v{e}d \v{C}esk\'e
           republiky, CZ-251 65 Ond\v{r}ejov, Czech Republic}

\date{Received 1 July 2003 / Accepted 24 November 2003}

\abstract{New numerical models of line-driven stellar winds of late
O stars are presented.
Statistical equilibrium (NLTE) equations of the most abundant elements
are solved.
Properly obtained occupation numbers are used to calculate consistent
radiative force and radiative heating terms.
Wind density, velocity and temperature are calculated as a solution
of model hydrodynamical equations.
Contrary to other published models we account for 
a
multicomponent wind
nature and do not simplify the calculation of the radiative force (e.g.
using force multipliers).
We discuss the convergence behaviour of our models.
The ability of our models to predict correct values of mass-loss rates
and terminal velocities of selected late O stars
(mainly giants and supergiants)
is demonstrated. 
The systematic difference between predicted and observed terminal
velocities reported in the literature has been removed.
Moreover, we found good agreement between the theoretical wind
momentum-luminosity relationship and the observed one
for Cyg OB2 supergiants.

    \keywords{stars: winds, outflows --
              stars:   mass-loss  --
              stars:  early-type --
             hydrodynamics
}
}

\maketitle

\section{Introduction}

NLTE line-driven stellar wind models 
have been
calculated by various authors
(cf. de Koter et al.  \cite{dekoter}, Hillier \& Miller \cite{hilmi},
Vink et al.  \cite{vikola}, Pauldrach et al. \cite{pahole}, Gr\"afener
et al.  \cite{grakoha}).
These sophisticated models cover a wide range of OB and WR star
parameters, and 
%
it may seem that
there is no need for any additional NLTE wind models.
However, we believe this is not the case.
Firstly, these models employ different simplifying assumptions which 
cannot
be necessarily fulfilled in all model stars.
For example, wind velocity is often calculated using either a simplified
$\beta$-velocity law (Vink et al. \cite{vikolamet}, Gr\"afener et al.
\cite{grakoha}) or using simplified line-force multipliers
(Pauldrach et al.  \cite{pahole}, Kudritzki \cite{kudmet}) of
Castor et al. (\cite{cak}) and Puls et al. (\cite{pusle}).
Some of these models (e.g., Vink et al. \cite{vikolamet}) do not allow
independent calculation of observed terminal velocities.
Moreover, temperature stratification is often neglected (cf. Kudritzki
\cite{kudmet}) assuming that the radiative force is not significantly
influenced by temperature (see Pauldrach \cite{pasam}) or calculated
using a simplified expression (Pauldrach et al. \cite{lepsipasam},
Vink et al. \cite{vikolabis}). 
Although the calculation of correct temperature stratification is
included 
in
modern wind codes (cf. Pauldrach et al. \cite{pahole}),
the latest published grid of NLTE calculations of wind temperature 
dates back to
the paper of Drew (\cite{drewmoc}), who used a relatively constrained
set of statistical equilibrium equations. Despite 
the abovementioned
enormous effort in 
%
wind model 
calculations,
many
unsolved problems 
remain
which may be connected either with the simplifying assumptions
mentioned above or with some others, e.g. neglect of wind
non-stationarity (cf. Runacres \& Owocki \cite{runow}),
wind 3D structure (Dessart \& Owocki \cite{des}),
or wind magnetic fields (ud-Doula \& Owocki \cite{asta}).
Among these unsolved problems 
is
the long-standing momentum
problem of WR stars (for the latest progress see Gr\"afener et al.
\cite{grakoha}) or problems with fitting of X-ray line profiles (Kramer
et al. \cite{kraco}).

Many of the simplifying assumptions employed in modern wind codes
are probably acceptable
for high density 
radiatively-driven
stellar winds
of O stars, but they may break down for low density stellar winds.
Whereas the bulk wind material consists of hydrogen and helium,
line-driven stellar winds are accelerated mainly by the absorption of
radiation in the resonance lines of trace elements
like C, N, O, or Fe.
For low-density stellar 
winds
this multicomponent nature may be important
for 
the
wind structure itself, influencing both temperature stratification
(Springmann \& Pauldrach \cite{treni}, Krti\v cka \& Kub\'at
\cite{kkii}, hereafter KKII) and  wind dynamics (Babel \cite{babela},
Porter \& Skouza \cite{reac}, KKII, Owocki \& Puls \cite{op}, Krti\v cka
\& Kub\'at \cite{kkiii}).
Clearly, such 
a
complicated situation 
cannot
be adequately described
by models with a 
temperature-independent
radiative force, or with a
radiative force {\em artificially} depending on 
%
wind temperature via
the thermal velocity as has been done, e.g., by KKII. Note
that the artificial dependence on the thermal velocity may be
removed using an alternative description of the radiative force by
Gayley (\cite{gayq}).

The situation is especially appealing now, when wind models
suitably describing the atmospheres of first generation stars have been
calculated (Kudritzki \cite{kudmet}), because some of these models are
also influenced by multicomponent effects (Krti\v cka et al.
\cite{gla}, see also Kudritzki \cite{kudmet}).
Moreover, predicted mass-loss rates of low-metallicity winds of some SMC
stars are more 
than
one magnitude higher than the observed ones (Bouret
et al. \cite{boula}).
These discrepancies may be connected with problems with source
function gradients (Owocki \& Puls \cite{owpu}).

Although satisfactory determination of mass-loss rates and terminal
velocities of O stars can be performed from observations in different
wavelength regions, the situation in the B star domain is not so clear.
Even for massive B stars there is a significant discrepancy between
observations in different wavelength regions (Vink et al.
\cite{vikola}).
Even worse, for most 
%
main-sequence B stars there is no trustworthy
determination of either mass-loss rates or terminal 
velocities, and,
moreover, it is not clear which B stars 
%
posses a stellar wind (see
Babel (\cite{babelb}) for the discussion of this problem).
This is a challenge for sophisticated wind models to help 
%
supplement
missing observational data.
There exist several subclasses of B stars whose peculiar behaviour may
be connected with their stellar wind.
This is epecially the case of Be stars, because the plausible
explanation of 
the
Be phenomenon is still missing (excluding some binaries,
for which a binary hypothesis of {\Kriz} \& Harmanec (\cite{kriha})
offers a trustworthy explanation).

Last but not least, there is an ambitious idea of the wind
momentum-luminosity relationship
(see Kudritzki \& Puls (\cite{kupul}) and references therein)
%
which may be used as an independent reliable method for determination of
stellar distances. Thus, several types of detailed
wind models are necessary 
%
to verify all of the assumptions of line-driven wind models.

Motivated by these considerations we offer a new independent type of
NLTE line-driven wind 
model
to obtain correct values
of radiative force, mass-loss 
rates
and terminal velocities in the case
of low-density stellar winds.

\section{Model assumptions}

Hydrodynamic model equations solved in this paper are essentially the
same as in KKII.
An interested reader will find more details about the solution of
multicomponent hydrodynamical equations in KKII and Krti\v{c}ka
(\cite{krt}).
Models presented here differ in the method of calculation of the
radiative
force and the
radiative heating/cooling term, and these new models assume
slightly
different boundary velocities.
In the following sections we 
summarize
our model equations.
The basic model assumptions are the following:
\begin{itemize}
\item We assume
a stationary
spherically-symmetric
flow.
\item
We solve the continuity equation, the momentum 
equation
and the energy
equation for each component of the flow, namely for absorbing ions,
nonabsorbing ions (hydrogen and helium), and electrons.
Nonabsorbing ions are ions for which the effect of a radiative
force may be neglected
(e.g., hydrogen and helium if
Population
I/II stars are considered).
%
\item Level occupation numbers of model ions are obtained from NLTE rate
equations with an inclusion of a superlevel concept (Anderson
\cite{limosuper}).
\item Radiation transfer is
split
into two parts.
Line radiative transfer is solved in a Sobolev approximation 
(Rybicki \& Hummer \cite{rybashumrem})
neglecting
continuum opacity sources, line overlaps, and multiple scattering.
Continuum radiative transfer is formally solved using the Feautrier
method with an inclusion of all free-free and bound-free transitions of
model ions, however neglecting line transitions.
Thus, we neglect 
line-blocking.
%
\item For the solution of NLTE rate equations together with the
radiative transfer equation we apply the Newton-Raphson method for the
radiative transfer in lines and lambda iterations in continuum.
The simplification of a lambda iteration may, of course, fail for
optically thick continua, like 
the
\ion{He}{ii} continuum
(however, see Sects.~\ref{resnlte} and \ref{chovani} for a discussion of
this issue).
\item Derived occupation numbers are used to calculate radiative force
in the hydrodynamic equations. 
\item 
The radiative
cooling/heating term is obtained by adding the
contributions from all free-free, bound-free and bound-bound transitions
of model ions.
For the calculation of this term we use the thermal balance of electrons
method (Kub\'at et al. \cite{kpp}).
\item The description of a transition region between 
a
quasi-static
stellar
atmosphere and supersonic stellar wind is simplified.
This region is important for the continuum formation (Pauldrach
\cite{pasam}).
This subsonic region was not included in our previous models since we
did not 
solve
the radiative transfer equation (see KKII).
To
obtain correct ionizing flux we included this subsonic
region 
in
our models, 
i.e.
the boundary velocity at the stellar
surface is set to some low value well below the sonic speed.
The particular value of the base velocity is selected in such a way that
lowering of this value by a factor of 10 does not change the model
structure (i.e. mass-loss rate and terminal velocity) 
by more than
10\%.
For most of the wind models this means that the boundary velocity is of
the order 
of
$10^{-3}$ of 
the
hydrogen thermal speed.
\item The calculation of models below and above the critical point
is split 
in
two parts, thus we neglect the influence of the radiation
emitted in the region above the critical point on the region below the
critical point. However, the influence of the subcritical
parts on the supercritical ones is taken into account correctly.
\item We neglect 
Gayley-Owocki heating (Gayley \& Owocki \cite{go})
which is not important for stars considered in this paper (see KKII).
\item We also neglect the X-ray radiation which may influence the
ionization balance.
\end{itemize}
These 
%
assumptions are discussed in more detail in the
Sect.~\ref{disko}.

\subsection{Equations of statistical equilibrium}

Our treatment of statistical equilibrium equations is very similar to
that of Pauldrach (\cite{pasam}).
We solve rate equations in the form of (Mihalas \cite{mihalas})
\begin{equation}
\label{nlte}
\sum_{j\neq i}N_j P_{ji}-N_i \sum_{j\neq i}P_{ij}=0,
\end{equation}
where $N_i$ is the number density of ions in state $i$ relative to the
total number density of given atom, and $P_{ij}$, $P_{ji}$ are rates of
all processes that transfer an atom from a given state $i$ to state $j$
and back, respectively.
They are sums of radiative and collisional rates (both
excitation/de-excitation
and ionization/recombination).
The system of rate equations \eqref{nlte} is closed by the equation of
normalisation of relative number densities for each model atom
(i.e. abundance equation)
\begin{equation}
\label{jednicka}
\sum_{i} N_i=1.
\end{equation}
This equation can be used instead 
of
any equation from the set
\eqref{nlte}.
However, it is computationally convenient to replace the rate equation
for the level with the highest relative number density.

%
We use the relative number density $N_i$ instead of the commonly used
number density to be consistent with the hydrodynamical part of the
code, because relative quantities are advantageous from the numerical
point of view.
Note that relative number densities were used for the solution of the
equations of statistical equilibrium by Lucy (\cite{eseiter}).

\subsubsection{Model atoms and rates}
\label{kapmodat}

Radiative excitation rates $R_{ij}$ are given by
\begin{equation}
\label{zarex}
N_i R_{ij} = N_i B_{ij} \bar J_{ij},
\end{equation}
where $\bar J_{ij} = \int_0^\infty \phi_{ij}(\nu) J_\nu \de \nu$
is profile-weighed line intensity and
\begin{equation}
B_{ij} = \frac{4\pi^2 e^2}{h\nu_{ij}m_\mathrm{e}c}f_{ij},
\end{equation}
where $\nu_{ij}$ is line frequency, $f_{ij}$ is corresponding oscillator
strength, $e$ is elementary charge and $m_\mathrm{e}$ is electron mass.

Similarly, the radiative 
de-excitation
rates $R_{ji}$ are
\begin{equation}
\label{zarde}
N_j R_{ji} = N_j (A_{ji}+B_{ji}\bar J_{ij}),
\end{equation}
where the remaining Einstein coefficients are given by
\begin{align}
A_{ji}&=\frac{2h\nu_{ij}^3}{c^2}B_{ji},
& B_{ji}&=\frac{g_i}{g_j} B_{ij},
\end{align}
and $g_i$, $g_j$ are statistical weights of involved levels.

Radiative ionization and recombination rates are given by
\begin{subequations}
\label{zarion}
\begin{align}
N_i R_{ik} &= 4\pi N_i \int_{\nu_i}^{\infty}
\frac{\alpha_{i,\nu}}{h\nu}J_\nu\,\de\nu,\\
N_k R_{ki} &=4\pi N_k \zav{\frac{N_i}{N_k}}^\ast
\int_{\nu_i}^{\infty}
\frac{\alpha_{i,\nu}}{h\nu}\hzav{\frac{2h\nu^3}{c^2}+J_\nu}
e^{-\frac{h\nu}{kT_\mathrm{e}}}
\,\de\nu,
\end{align}
\end{subequations}
where $\alpha_{i,\nu}$ is the corresponding photoionization
cross-section, $J_\nu$ is the mean
continuum intensity (line absorption
spectrum is neglected in calculations of bound-free rates), $\nu_i$ is
the ionization frequency of level~$i$, $T_\mathrm{e}$ is the electron
temperature, and an asterisk denotes LTE values.

Upward ($C_{ij}$) and downward ($C_{ji}$)
collisional rates (for both excitation and
ionization) are calculated using the expressions
\begin{subequations}
\label{srazky}
\begin{align}
N_i C_{ij} &= N_i n_\mathrm{e} \Omega_{ij} (T_\mathrm{e}),\\
N_j C_{ji} &= N_j n_\mathrm{e} \zav{\frac{N_i}{N_j}}^* \Omega_{ij} (T_\mathrm{e}),
\end{align}
\end{subequations}
where $\Omega_{ij} (T_\mathrm{e})$ is the integrated collisional
cross-section (collision strength) and $n_\mathrm{e}$ is the electron
number density. Finally, rate coefficients $P_{ij}$ for both
excitation and ionization are sums of radiative and collisional terms
\begin{equation}
\label{pijsou}
P_{ij}=R_{ij}+C_{ij}.
\end{equation}

\begin{table}
\caption{Atoms and their ionization stages included 
in
the NLTE
calculations.
In this table,
level
means either an individual level or a
set of levels merged into a superlevel.
}
\label{prvky}
\centering
\begin{tabular}{lrlrlr}
\hline\hline
Ion & Levels & Ion & Levels & Ion & Levels \\
\hline
 \ion{H}{i}   &   9  &  \ion{Ne}{iv} &  12 &  \ion{S}{iii}  &  10  \\
 \ion{H}{ii}  &   1  &  \ion{Ne}{v}  &  17 &  \ion{S}{iv}   &  18  \\
 \ion{He}{i}  &  14  &  \ion{Ne}{vi} &   1 &  \ion{S}{v}    &  14  \\
 \ion{He}{ii} &  14  &  \ion{Na}{ii} &  13 &  \ion{S}{vi}   &   1  \\
 \ion{He}{iii}&   1  &  \ion{Na}{iii}&  14 &  \ion{Ar}{iii} &  25  \\
 \ion{C}{ii}  &  14  &  \ion{Na}{iv} &  18 &  \ion{Ar}{iv}  &  19  \\
 \ion{C}{iii} &  23  &  \ion{Na}{v}  &  16 &  \ion{Ar}{v}   &  16  \\
 \ion{C}{iv}  &  25  &  \ion{Na}{vi} &   1 &  \ion{Ar}{vi}  &   1  \\
 \ion{C}{v}   &   1  &  \ion{Mg}{iii}&  14 &  \ion{Ca}{ii}  &  16  \\
 \ion{N}{ii}  &  14  &  \ion{Mg}{iv} &  14 &  \ion{Ca}{iii} &  14  \\
 \ion{N}{iii} &  32  &  \ion{Mg}{v}  &  13 &  \ion{Ca}{iv}  &  20  \\
 \ion{N}{iv}  &  23  &  \ion{Mg}{vi} &   1 &  \ion{Ca}{v}   &  22  \\
 \ion{N}{v}   &  13  &  \ion{Al}{ii} &  16 &  \ion{Ca}{vi}  &   1  \\
 \ion{N}{vi}  &   1  &  \ion{Al}{iii}&  14 &  \ion{Fe}{iii} &  29  \\
 \ion{O}{ii}  &  50  &  \ion{Al}{iv} &  14 &  \ion{Fe}{iv}  &  32  \\
 \ion{O}{iii} &  29  &  \ion{Al}{v}  &  16 &  \ion{Fe}{v}   &  30  \\
 \ion{O}{iv}  &  39  &  \ion{Al}{vi} &   1 &  \ion{Fe}{vi}  &  27  \\
 \ion{O}{v}   &  14  &  \ion{Si}{ii} &  12 &  \ion{Fe}{vii} &   1  \\
 \ion{O}{vi}  &  20  &  \ion{Si}{iii}&  12 &  \ion{Ni}{iii} &  36  \\
 \ion{O}{vii} &   1  &  \ion{Si}{iv} &  13 &  \ion{Ni}{iv}  &  38  \\
 \ion{Ne}{ii} &  15  &  \ion{Si}{v}  &   1 &  \ion{Ni}{v}   &  48  \\
 \ion{Ne}{iii}&  14  &  \ion{S}{ii}  &  14 &  \ion{Ni}{vi}  &   1  \\
\hline
\end{tabular}
\end{table}

We solve NLTE rate equations \eqref{nlte} for each of the elements
listed in Table~\ref{prvky} separately.
This allows us (in addition to a fixed temperature) to fix the number of
free electrons during the solution of the equations of statistical
equilibrium.
This step greatly reduces the computational needs, since instead of one
large rate matrix we have to deal with a moderate number of smaller
matrices for the elements considered.

The model atoms are predominantly taken from TLUSTY models
(Hubeny \cite{tlusty}, Hubeny \& Lanz \cite{hublaj}, Hubeny \& Lanz
\cite{hublad}, Lanz \& Hubeny \cite{lahub}).
The set of model atoms is slightly modified for the case of hot star
wind modelling.
For this purpose we used data from the Opacity Project (Seaton
\cite{top}, Luo \& Pradhan \cite{top1}, Sawey \& Berrington
\cite{savej}, Seaton et al. \cite{topt}, Butler et al. \cite{bumez},
Nahar \& Pradhan \cite{napra}) and from the Iron Project (Hummer et al.
\cite{zel0}, Bautista \cite{zel6}, Nahar \& Pradhan \cite{zel2},
Zhang \cite{zel1}, Bautista \& Pradhan \cite{zel5}, Zhang \& Pradhan
\cite{zel4}, Chen \& Pradhan \cite{zel3}).

We use 
a
similar strategy for the preparation of model atoms that enter
the calculations as was already applied in TLUSTY.
With 
the
exception of \element{Fe} and \element{Ni}, levels with low
energies (relative to the ground level of a given ionization stage) are
treated individually (neglecting the hyperfine splitting), whereas
higher levels with similar properties are merged into superlevels.
For \element{Fe} and \element{Ni} all levels (except the ground ones)
are merged into superlevels.
More details on the construction of model atoms
and atomic data used are given in Appendix \ref{atdata}.

Procedures for the calculation of $\Omega_{ij} (T_\mathrm{e})$ are taken
from the TLUSTY code.
The partition
function 
approximations
are mostly from Smith \& Dworetsky
(\cite{smid}) and Smith (\cite{smith}).
Moreover, 
the
partition function 
approximations
taken from the TLUSTY code
are applied for \element{C}, \element{N}, \element{O}, \ion{Fe}{vi},
and \ion{Fe}{vii}.

\subsubsection{Superlevels}

Compared to the pioneering work of Anderson (\cite{limosuper}), we
use a simplified merging of individual levels into superlevels.
Given the sublevels $l$ with energies $E_{il}$, relative number
densities $N_{il}$, and statistical weights $g_{il}$, the total
population of superlevel is given by
\begin{equation}
N_i=\sum_l N_{il},
\end{equation}
and 
the
energy of superlevel is 
\begin{equation}
E_i=\frac{\sum_l g_{il} E_{il}}{\sum_l g_{il}}.
\end{equation}
Thus, we effectively neglect the
$\exp(-E_{il}/kT_\mathrm{e})$ term.
Similarly, the photoionization cross section of superlevel
$\alpha_{i,\nu}$ is calculated by
\begin{equation}
\alpha_{i,\nu}=\frac{\sum_l g_{il} \alpha_{il,\nu}} {\sum_l g_{il}},
\end{equation}
and the relative number density of a sublevel of a given superlevel is
approximated as
\begin{equation}
\label{nijsup}
N_{il}=\frac{g_{il}N_i}{\sum_l g_{il}}.
\end{equation}
Numerical 
tests
showed that the neglect of the
$\exp(-E_{il}/kT_\mathrm{e})$
term 
does not have
a significant effect on the calculated radiative force. 

\subsection{Radiative transfer}

Radiative flux at the base of the stellar wind $H_c$ is taken from
\element{H}-\element{He} spherically symmetric NLTE model atmospheres of
Kub\'at (\cite{kub}).
The line transfer equation and continuum transfer equation are solved
separately.

\subsubsection{Line radiative transfer}

The averaged line intensity is calculated using the Sobolev
approximation (Rybicki \& Hummer \cite{rybashumrem})
\begin{equation}
\label{prenoscar}
\bar J_{ij}=(1-\beta)S_{ij} + \beta_c I_c,
\end{equation}
where escape probability functions $\beta$ and $\beta_c$ are given by
integrals
\begin{align}
\label{betac}
\beta_c&=\frac{1}{2}\int_{\mu_*}^{1}d\mu\frac{1-e^{-\tau_\mu}}
{\tau_\mu},\\
\label{beta}
\beta&=\frac{1}{2}\int_{-1}^{1}d\mu\frac{1-e^{-\tau_\mu}}{\tau_\mu},
\end{align}
$\mu_*=\zav{1-R_*^2/r^2}^{1/2}$,
and $I_c=4H_c$.
The Sobolev optical depth $\tau_\mu$ is given by (Castor \cite{cassob},
Rybicki \& Hummer \cite{rybashumrem})
\begin{equation}
\label{tau}
\tau_\mu=\frac{\pi e^2}{m_\mathrm{e}\nu_{ij}}
\zav{\frac{n_i}{g_i}-\frac{n_j}{g_j}} g_if_{ij}
\frac{r}{\io\vr\zav{1+\sigma\mu^2}},
\end{equation}
where
$\io\vr$ is the radial velocity of absorbing ions,
$n_i$, $n_j$ are the number densities of individual states, 
\begin{equation}
\label{ni}
n_i=N_iz_{\rm atom}n_\mathrm{H}
\end{equation}
(similarly for $n_j$),
where
$z_{\rm atom}=n_{\rm atom}/n_{\rm H}$
is 
the
number density of a given atom relative to the
hydrogen number density $n_\mathrm{H}$, 
\begin{equation}
n_\mathrm{H}=
\frac{\rho_\mathrm{p}}{m_\mathrm{H}+z_\mathrm{He}m_\mathrm{He}},
\end{equation}
where $\rho_\mathrm{p}$ is the density of a passive component (hydrogen
and helium).
The variable $\sigma$ was introduced by Castor (\cite{cassob}),
\begin{equation}\label{sigma}
\sigma=\frac{\de\ln \io\vr}{\de\ln r} -1.
\end{equation}
A suitable
method for the calculation of escape probability functions
$\beta$ and $\beta_c$ was described by KKII.
Finally, the line source function is
\begin{equation}
\label{zdrojcar}
S_{ij}=\frac{N_j A_{ji}}{N_i B_{ij}-N_j B_{ji}}.
\end{equation}

\subsubsection{Continuum radiative transfer}

For the calculation of the continuum intensity we use a procedure for
the solution of the transfer equation based on the Feautrier method in
the spherical coordinates (for a description, see Mihalas \& Hummer
\cite{sphermod} or Kub\'at \cite{dis}).
However, for the solution of the obtained tridiagonal system of
equations we use 
the
numerical package LAPACK
({\tt http://www.cs.colorado.edu/\~{}lapack},
Anderson et al. \cite{lapack}) instead of the classical Gaussian scheme.
All bound-free transitions of explicit ions, free-free transitions of
hydrogen and helium and light scattering due to free electrons
contribute to the source function.
The radiative transfer equation is solved only above the hydrogen
ionization edge.
For frequencies lower than the hydrogen ionization limit we use an
optically thin approximation $J_\nu(r) = 4 W(r) H_c$, where
$W(r)=\frac{1}{2}\zav{1-\mu_*}$ is the dilution factor.

\subsection{Solution of NLTE equations}
\label{resnlte}

For the solution of the NLTE rate equations \eqref{nlte},
\eqref{jednicka} together with the radiative transfer equation we use
a combination of lambda iterations and of a complete linearization.
For 
the
continuum transfer we use simple lambda iterations.
Although the stellar wind is optically thick near to the star 
below the
\ion{He}{ii} ionization limit, these iterations converge relatively well
(see also Sect.~\ref{chovani}).
In lines, we use complete linearization of the radiative transfer
equation in a simple Sobolev approximation (neglecting continuum
radiation, see Eq.\,(\ref{prenoscar})).
The line intensity $\bar J_{ij}$ depends only on the level populations
of the given atom for this case.
Thus, the statistical equilibrium equations for different atoms are
independent (for a {\em fixed}
hydrodynamical structure).
Hence, 
the
following scheme may be obtained from statistical equilibrium
equations \eqref{nlte}, \eqref{jednicka} for the calculation of the
correction $\delta N_i$ of relative number densities in the given
iteration step
\begin{multline}
\label{nlteskut}
\sum_{j\neq i}\hzav{N_j\frac{\partial P_{ji}}{\partial N_i}\delta N_i+
      \zav{P_{ji}+ N_j\frac{\partial P_{ji}}{\partial N_j}} \delta N_j}-
\\*
     -\sum_{j\neq i}\hzav{\zav{P_{ij}+
      N_i\frac{\partial P_{ij}}{\partial N_i}}\delta N_i+
      N_i\frac{\partial P_{ij}}{\partial N_j} \delta N_j}=0, 
\end{multline}
where the terms ${\partial P_{ij}}/{\partial N_j}$ are calculated using
Eqs.~\eqref{prenoscar} -- \eqref{zdrojcar}. 
For details see Appendix \ref{nlteres}.
NLTE equations \eqref{nlteskut} are solved for each element separately.
Note that Eq.~\eqref{nlteskut}
is 
%
an application of a Newton-Raphson method to
statistical equilibrium equations
\eqref{nlte}.
For the solution of 
the
system of NLTE equations \eqref{nlteskut} we use
the numerical package LAPACK.
Note that we can again take full advantage of LAPACK subroutines
designed
for general band matrices because
our
NLTE rate 
equation matrices have
a band structure.
%

\subsection{Derivatives of occupation numbers}

The hydrodynamic equations are solved using a Newton-Raphson method.
To
obtain 
a
well converged model we have to know the explicit
values of derivatives of the radiative force and of the radiative
heating/cooling term with respect to fluid variables (i.e. densities,
temperatures, etc.).
However, because the radiative force and the radiative heating/cooling
term depend on level populations, we have to calculate derivatives of
level populations with respect to fluid variables.
These can be calculated 
in
the following way.
For clarity, we rewrite NLTE rate equations \eqref{nlte} in a symbolic
form
\begin{multline}
\label{nltesym}
\sum_{j\neq i}P_{ji}
(\vec{N},J_\nu(\vec{N},{\vec{h}}),\vec{h}) 
N_j-\\*
-N_i\sum_{j\neq i} P_{ij}
(\vec{N},J_\nu(\vec{N},\vec{h}),\vec{h})
=0,
\end{multline}
where we denoted all explicit dependencies of rate equations.
The elements of vector $\vec{N}$ are level populations $N_i$, the
elements of vector $\vec{h}$ are fluid variables $\vec{h}=(n_\mathrm{e},
\rho_\mathrm{p},T_\mathrm{e},\io\vr)$, and the mean continuum
intensity $J_\nu$ 
depend
on $\vec{h}$ and $\vec{N}$ via the radiative
transfer equation. Clearly, it is not feasible
to obtain correct explicit values of $\partial N_i / \partial h$ 
(where $h$ denotes 
elements
of vector $\vec{h}$),
since the level populations depend on the solution of radiative transfer
equation in continuum. However, it is possible to obtain
approximate values of these derivatives
neglecting the explicit dependence of rates $P_{ij}$ on the continuum
intensity. 
Deriving
Eq.~\eqref{nltesym} with respect to 
$h$ and neglecting the dependence on $J_\nu(\vec{N},\vec{h})$
we obtain a system of equations for $\partial N_i / \partial h$
\begin{multline}
\label{nlteder}
\sum_{j\neq i}\left[\frac{\partial P_{ji}}{\partial h} N_j+
    \frac{\partial P_{ji}}{\partial N_i} \frac{\partial N_i}{\partial h} N_j+
    \zav{\frac{\partial P_{ji}}{\partial N_j} 
N_j+P_{ji}} \frac{\partial N_j}{\partial h}\right]-\\*
\sum_{j\neq i}\left[\frac{\partial P_{ij}}{\partial h} N_i+
    \frac{\partial P_{ij}}{\partial N_j} \frac{\partial N_j}{\partial h}N_i+
    \zav{\frac{\partial P_{ij}}{\partial N_i} 
N_i+P_{ij}} \frac{\partial N_i}{\partial h}
\right]\\*=0.
\end{multline}
All calculated derivatives $\partial N_i / \partial h$ are inserted into
the Newton-Raphson matrix of hydrodynamic equations (corresponding to
the radiative force, radiative heating/cooling term and ionization
charge, see KKII and Krti\v{c}ka \cite{krtdis}).

Note that the system of equations for derivatives \eqref{nlteder} has
the same form as the system of NLTE rate equations \eqref{nlteskut},
however with a different right-hand side.
Thus, because 
the LAPACK package
uses an LU decomposition for the solution
of the system of linear equations, the values of
$\partial N_i / \partial h$ can be calculated using data from the
previous solution of NLTE equations (\ref{nlteskut}).

This method is especially useful in the case when the intensity of
radiation in 
the
continuum does not depend on local properties (i.e.,
for an optically thin continuum, when it is given by the
intensity emerging from the underlying photosphere).
From the mathematical point of view, this method splits the Jacobi
matrix into several parts and it is equivalent to a linearization of the
whole system of model equations (i.e. hydrodynamics, NLTE equations and
radiative transfer).
However, from the numerical point of view, it saves a huge amount of
computer time because it is not necessary to invert large matrices.
A similar
method has been used, e.g., by Anderson (\cite{mfmg}), and
Kub\'at (\cite{M1,M3}) referred to it as to the implicit linearization
of the {\em b}-factors.

\subsection{Calculation of the radiative force}

The radiative force is calculated using the Sobolev approximation
directly as the sum of contributions of individual lines after Castor
(\cite{cassob}),
\begin{equation}
\label{zarzrych}
\io g^{\rad}=\frac{8\pi}{\io\rho c^2}\frac{\vr}{r}
      \sum_{\mathrm{lines}}\nu H_c\int_{\mu_c}^{1}\mu\zav{1+\sigma\mu^2}
      \zav{1-e^{-\tau_\mu}},
\end{equation}
where $\tau_\mu$ is given by 
Eq.~\eqref{tau}.
Occupation numbers $N_i$ calculated during the NLTE iteration step are
used for a calculation of the radiative force.  
Note that in 
Eq.~\eqref{tau} number densities of {\em individual}
levels are used instead of the total number density of a superlevel.
These number densities may be calculated from 
Eq.~\eqref{nijsup} for
those levels which are explicitly included 
in
the NLTE rate equations.
However, non-explicit levels, i.e., some of the higher excited levels
or excited levels of the highest ion of an atom, are also included 
in
the calculation of the radiative force.
For these non-explicit levels we use either 
the
Boltzmann equilibrium at
the wind temperature in the case of metastable levels or 
the
radiative
excitation equilibrium of Abbott \& Lucy (\cite{abblu}) for other
levels.
The influence of these non-explicit levels on the radiative force is
very small, however.

The term $\partial \io g^{\rad} / \partial\zav{\de\io\vr/\de r}$ is
calculated explicitly after \eqref{zarzrych} where the variable
$\sigma$ depends on a velocity gradient via \eqref{sigma}.
This term
is included 
in
the critical point condition (generalised CAK
condition, KKII, Eq.~(57) therein).
Derivatives of number densities calculated by means of Eq.~\eqref{nlteder}
are used to obtain derivatives of the radiative force with respect to
fluid variables (densities, velocities, and temperatures) for the
Newton-Raphson iterations.

Oscillator strengths necessary for the calculation of the radiative
force are extracted from the VALD database (Piskunov et al.
\cite{vald1}, Kupka et al.  \cite{vald2}).
This set of data consists of about 180\,000 line transitions in the
wavelength interval $250$ -- $10\,000\,$\AA.

\subsection{Radiative cooling and heating}

The radiative
cooling and/or heating term is calculated using the thermal
balance of electrons (Kub\'at et al. \cite{kpp}).
For the calculation of this term we include all the explicit radiative
bound-free and collisional (both bound-bound and bound-free)
transitions and free-free transitions of
hydrogen and helium.
Again, in order to obtain a well-converged model, derivatives of number
densities calculated by means of 
Eq.~\eqref{nlteder} are used to
obtain derivatives of the radiative cooling and/or heating terms with
respect to fluid variables. The 
calculated derivatives of 
the
radiative 
cooling/heating
term are inserted
into the Jacobi matrix of the Newton-Raphson iteration of the model
hydrodynamic structure.
Detailed expressions for the derivatives of heating and cooling rates
are presented in Appendix\,\ref{heacoo}.

\subsection{Electron density}

The charge of 
the
passive component is calculated from ionization
fractions of both hydrogen and helium.
This enables 
us
to calculate 
the
correct value of the electron density, which
is important for the proper solution of NLTE rate equations.

\subsection{Iteration scheme for the model calculation}
\label{jakkonverkap}

\begin{figure*}
\centering
\resizebox{0.9\hsize}{!}{\includegraphics{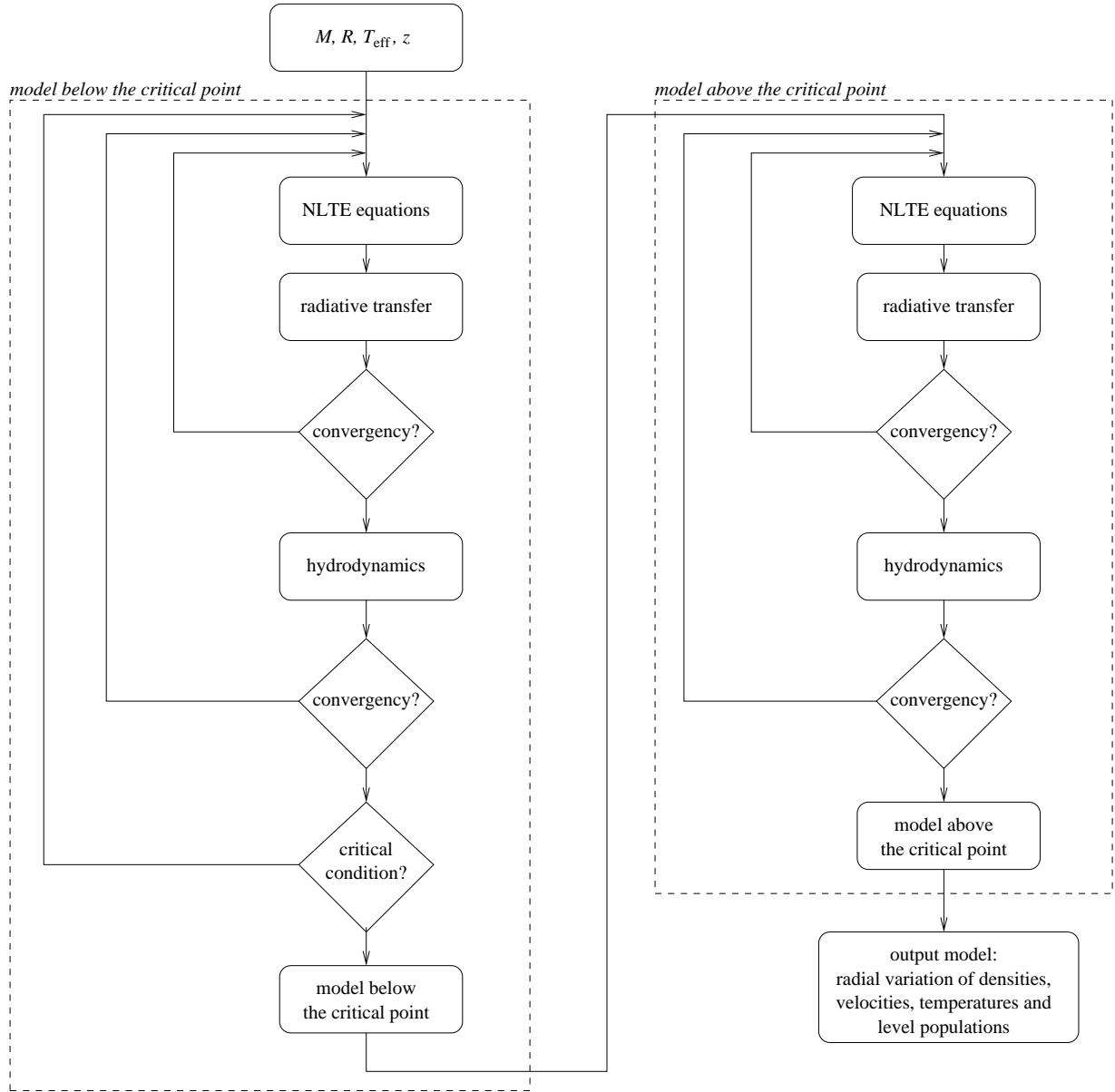}}
\caption{Iteration scheme for the model calculation}
\label{jakpocit}
\end{figure*}

The initial estimate of fluid variables for the Newton-Raphson
iterations (namely the so-called beta-law for velocities and density
obtained from 
the
continuity equation) was described by KKII.
For the initial value of the intensity we use an optically thin
approximation, i.e., the radiation field is defined by the lower
boundary condition at the beginning of our calculation.

The computational
procedure involves two similar basic steps, namely
solution below and above the wind critical point.
This splitting into two steps is necessary and is related to the
critical point condition. In the first step the value of the base
density, which allows the
model to smoothly pass the critical point, is obtained.
Afterwards, the model above this critical point is calculated.

The calculation of a given model below the critical point consists of
three nested iteration cycles (see Fig.~\ref{jakpocit}).
In the innermost iteration cycle the NLTE rate equations \eqref{nlte}
together with the radiative transfer equation both in lines
\eqref{prenoscar} and in continuum are solved.
For this innermost iteration cycle the fluid variables (i.e. densities,
velocities and temperatures of individual wind components) are kept
fixed.

In the middle cycle we iterate 
the
hydrodynamical structure of the model
for fixed level populations.
This iteration cycle is essentially the same as the iteration cycle of
three-component nonisothermal wind models described by KKII.
To save computing time, for 
large relative changes of fluid variables (greater than
$10^{-2}$) we perform only 3 iterations of NLTE rate equations, but when
the relative changes of fluid variables are lower,
and we may expect that we are closer to the correct solution,
we perform more iterations of NLTE rate equations (up to 50).
We assume that model is well converged if the relative changes of
{\em both}
fluid variables and occupation numbers are lower than $10^{-5}$.

In the outermost iteration cycle we solve for the value of the base
density for which a model smoothly passes through the critical point.
Again, these iterations were already described by KKII. For the 
base density (which determines the value of mass-loss rate
$\de{\eu M}/\de t$) 
an
accuracy of $1\%$ is sufficient.

After the base wind density is known, we calculate wind model above the
critical point by a sequence of iterations of fluid variables.
Each of these iterations is preceded by the set of NLTE iterations.
Boundary conditions of the model above the critical point are specified
at the critical point and are taken from the model below the critical
point.
The model below the critical point is fixed during the solution above
the critical point.

Splitting of a wind solution into two parts may
cause 
%
problems since the radiation emitted from 
the
outer region
(i.e. above the CAK point) may influence 
the
inner region.
However, test calculations where we changed 
the
outer boundary of inner
solution showed that this is not the case, i.e. that 
the
inner wind solution
depends only marginally on the outer solution.
The reason for such a behaviour is probably the strong dependence of
the
wind continuum optical depth on 
frequency.
Calculations show
that the stellar wind above the CAK point is 
essentially
optically thin for continuum radiation
with frequencies below the \ion{He}{ii} ionization limit, thus radiation
is not absorbed in 
the
continuum above the CAK point and does
not influence the stellar wind below the CAK point.
On the other hand, stellar wind above the CAK point is optically very
thick for radiation with frequencies above the \ion{He}{ii} ionization
limit.
Again, due to this large optical thickness the radiation emitted in the
regions above the CAK point does not influence the inner stellar wind.
The
influence of lines above the 
critical point on the subcritical region has been also neglected.

\section{Convergence behaviour of models}
\label{chovani}

\begin{figure*}
\resizebox{0.5\hsize}{!}{\includegraphics{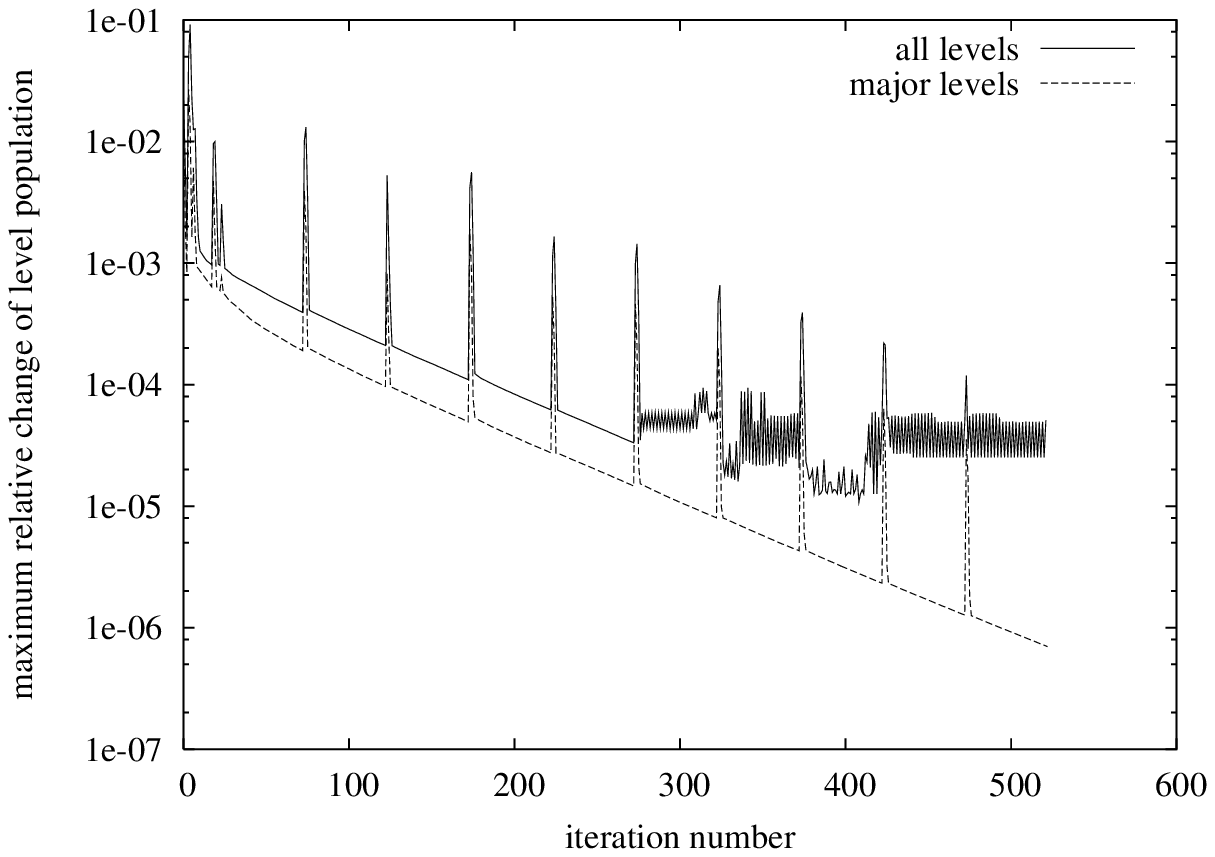}}
\resizebox{0.5\hsize}{!}{\includegraphics{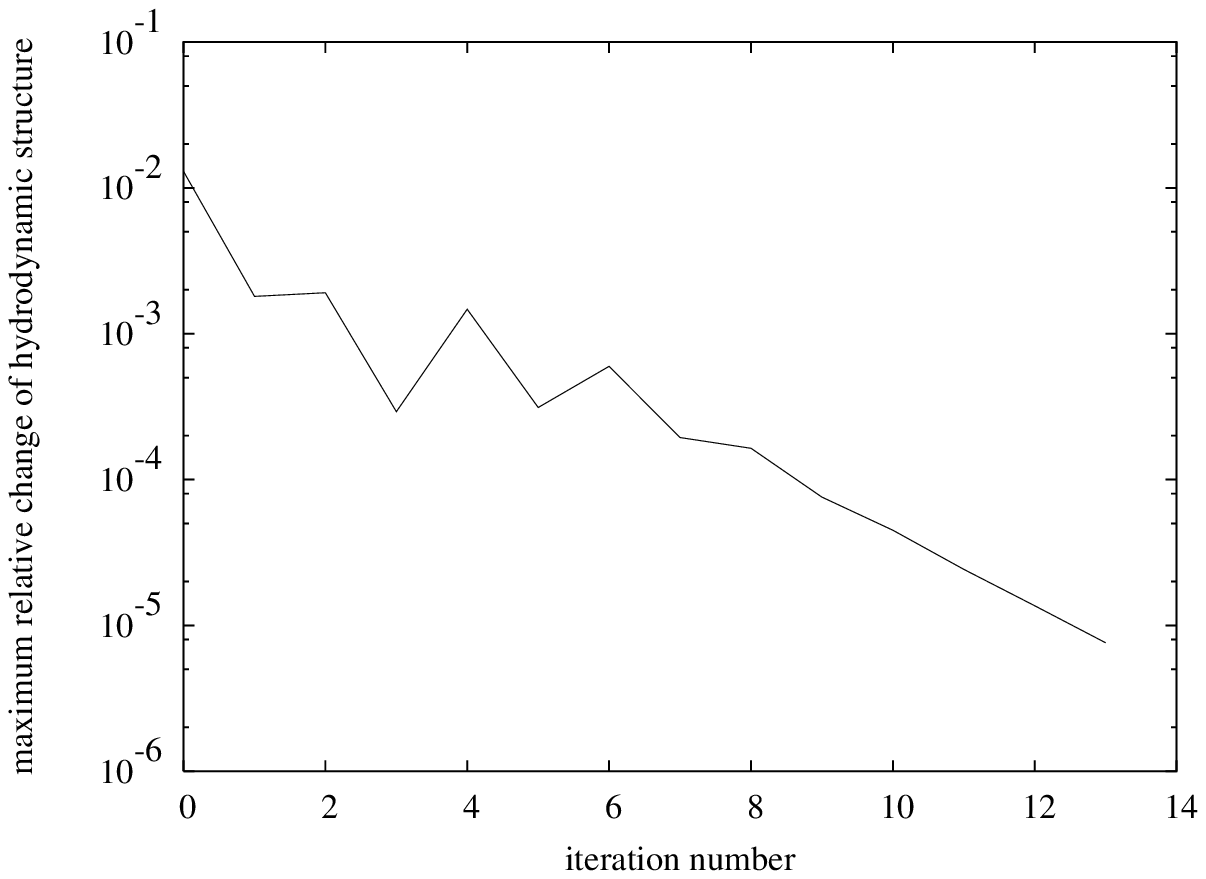}}
\resizebox{0.5\hsize}{!}{\includegraphics{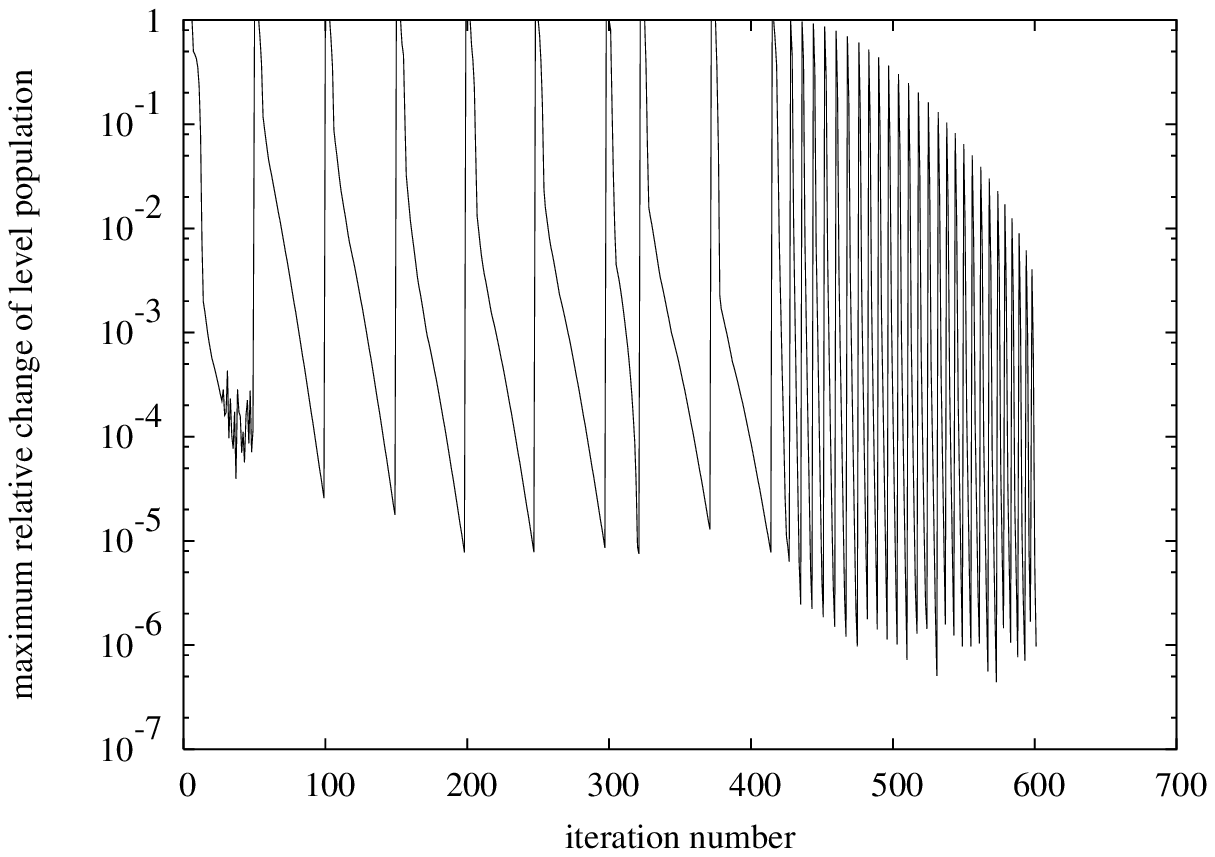}}
\resizebox{0.5\hsize}{!}{\includegraphics{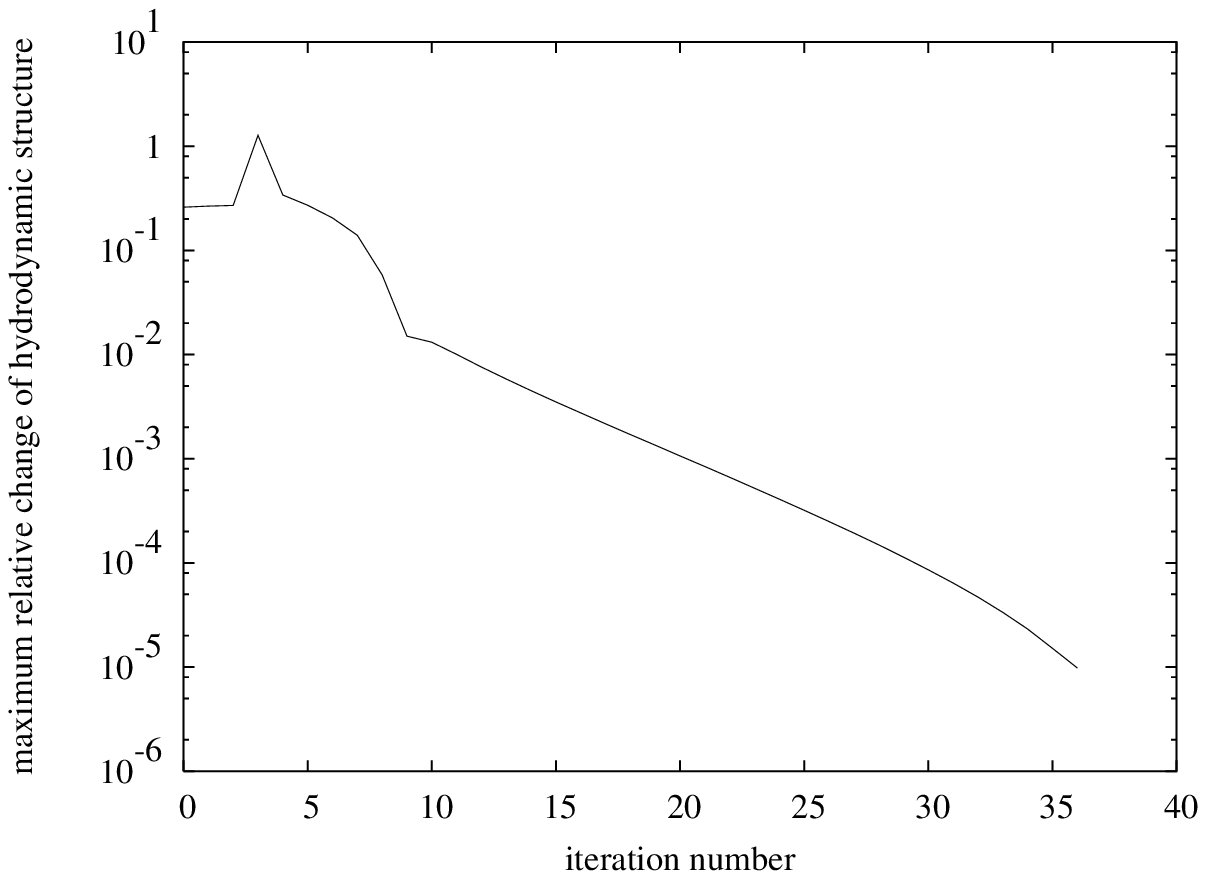}}
\caption[]{The convergence of the
models below (upper panels) and above (lower panels) the critical point.
{\em Upper left panel:} The convergence of level populations
below the critical point.
Note that NLTE iterations in consecutive hydrodynamical iterations are
combined
in this plot.
This explains 
the
relatively large number of 
iterations
and peaks 
that
occur after the hydrodynamical iteration step.
The oscillating behaviour for the relative change of order $10^{-5}$
is due to the weakly populated levels, because there are no oscillations
in the plot of convergence of relative change of major levels (i.e.
levels with $N_i>0.01$).  
{\em Upper right panel:} The convergence of hydrodynamical structure
(maximum relative change of all hydrodynamical variables,
i.e. densities, velocities, temperatures and charges of individual wind
components) of the model below the critical point.
The slightly oscillating behaviour around the relative change $10^{-3}$
is caused by the change of maximum allowed number of NLTE iteration
steps (see the text).
{\em Lower left panel:} The convergence of level populations of the
model above the critical point.
Again, NLTE iterations in different hydrodynamical iterations are put
together and peaks occur after the hydrodynamical iteration step.
Note the fast convergence of NLTE population between successive
hydrodynamical iterations.
{\em Lower right panel:} The convergence of the hydrodynamical
structure of the model above the critical point.}
\label{konverob}
\end{figure*}

As an example we study the convergence behaviour of a model
corresponding to 
$\alpha$~Cam (\object{HD\,30614})
with parameters given in Table \ref{ohvezpar}. In 
Fig.~\ref{konverob} we plotted relative change of population
numbers, and hydrodynamical variables (i.e. densities, velocities,
temperatures and charges of individual wind components) for the
model below the critical point and for the model above the critical
point.

First we discuss the convergence behaviour of the model below the
critical point.
For the plot of relative change of level populations (upper left panel
of Fig.~\ref{konverob}) we 
combine
NLTE iterations performed during
individual hydrodynamical iteration steps
for fixed base wind density (corresponding to the final CAK solution).
Thus, in this plot
the
peaks corresponding to the hydrodynamical iteration steps occur.
They are caused by a reaction of NLTE populations to a significant
change of the fluid variables after the calculation of the new estimate
of the hydrodynamic structure.
For the relative change
of
about $10^{-5}$ relative changes oscillate.
However, if we plot only relative changes of populations of levels with
largest occupation numbers (levels with $N_i>0.01$), no oscillations
appear.
Thus, we conclude that the oscillatory behaviour is caused by
weakly populated levels and 
that
their influence on the convergence behaviour
of the hydrodynamical structure is negligible
(upper right panel of Fig.~\ref{konverob}).
This oscillatory behaviour is not a typical feature of
convergence of our models.

The plot of hydrodynamical structure convergence
(upper right panel of Fig.~\ref{konverob}) shows very fast
convergence which is slightly slowed down by the oscillatory behaviour
around the relative change $10^{-3}$.
This is caused by the change of 
the
maximum allowed number of NLTE iteration
steps (see Sect.~\ref{jakkonverkap}).
Although due to this change the convergence is slightly slowed down, the
total computing time is 
less
because we do not
lose
time with calculation of
precise occupation numbers for hydrodynamical structure which is far
from the converged one. Finally, note that the relatively small initial
relative change of fluid variables
(of order $10^{-2}$) of 
the
hydrodynamical structure is
caused by the way 
%
the critical point is calculated,
ie. that models with similar base densities are calculated.
Thus, the initial change of fluid variables of the final model below the
critical point is very small.
The convergence behaviour of previous models (i.e. models which do not
correspond to the critical solution) is similar, however the initial
change of fluid variables is higher.

The convergence behaviour of the model above the critical point (lower
panels of Fig.~\ref{konverob}) is similar to that below the critical
point.
The convergence of occupation numbers (for a given hydrodynamical
structure) is faster because the stellar wind is optically thinner (note
that this fast convergence of NLTE iterations combined with larger
number of iterations of hydrodynamic structure yields something like a
"forest" of iterations in the plot, which is caused by a large change of
occupation numbers after a new hydrodynamic structure was calculated.
The convergence of a hydrodynamical structure (lower right panel of
Fig.~\ref{konverob}) is however slower than below the critical point.
This is partially caused by lower maximum change of wind parameters
allowed during Newton-Raphson iteration step. This 
constraint of the
maximum change of wind parameters is necessary for
the stability of the solution (see Krti\v cka \cite{krt}).

The convergence
of 
the
model temperature is very fast in the case of a model below the
critical point. The relative change of temperature
is typically ten times lower than the relative change of the rest of
the
hydrodynamical variables in this case. On the other hand, the convergence
behaviour of a model above the critical point is dominated by the
temperatures since the relative
changes of temperature are the highest.

\section{Comparison of calculated and observed quantities}

\begin{table*}[hbt]
\begin{center}
\caption{Stellar and wind parameters of selected O~stars from our test
sample.
Stellar parameters (columns ${\eu M}$, $R_{*}$, $\Teff$) were adopted
from Lamers et al. (\cite{lsl}).
Observed mass loss rates were taken from 
[1] -- Leitherer (\cite{lei}),
[2] -- Scuderi et al. (\cite{scupa}),
[3] -- Lamers \& Leitherer (\cite{lamlei}),
[4] -- Puls et al. (\cite{pulmoc}),
[5] -- Crowther \& Bohannan (\cite{croboh}),
[6] -- Scuderi et al. (\cite{scupamoc}),
[7] -- Lamers et al. \cite{lamoc},
[8] -- BG (see the text),
predicted values of
$\dot{\eu M}$ and $v_{\infty}$ are determined using our code.
Note
that terminal velocities calculated by KKII were obtained using
simplified force multipliers after Abbott (\cite{abpar}).}
\label{ohvezpar}
\begin{tabular}{rrccrcccccccc}
\hline
\hline
\multicolumn{1}{c}{Star} & \multicolumn{1}{c}{HD} & Sp.
&\multicolumn{3}{c}{Stellar parameters} & 
\multicolumn{4}{c}{Mass loss rates $\dot{\eu M}$} &
\multicolumn{3}{c}{Terminal velocities $v_\infty$} \\
& \multicolumn{1}{c}{number}& type
&${\eu M}$ & $R_{*}$ & $\Teff$ & 
H$\alpha$ & radio & UV & model & LSL, BG  & KKII & model
\\ & & & $[{\eu M}_{\odot}]$ & $[R_{\odot}]$ & $[\mathrm{K}] $ & 
\multicolumn{4}{c}{$[10^{-6}\,{\eu M}_{\odot}\,\mathrm{yr}^{-1}]$} &
\multicolumn{3}{c}{$[\mathrm{km}\,\mathrm{s}^{-1}]$}\\
\hline
$\alpha$  Cam&$ 30614$ &  O9.5 Ia & $43.0$ & $27.6$ & $30900$ & 
 $5.2\ct$; $2.9\se$ & $2.5\se$ & & $2.4$ & $1500\pm200$ & $1450$ & $1470$ \\
&$ 34656$ &  O7 II & $30.0$ & $ 9.9$ & $38100$ & 
   $0.49\je$ & & & $0.54$ & $2100\pm100$ & $2590$ & $2410$ \\
$\lambda$  Ori A&$ 36861$ & O8 III & $30.0$ & $12.3$ & $36000$ & 
   $0.81\sm$ & & & $0.66$ &$2200\pm300$ & $2230$ & $2260$ \\
 V973 Sco &$151804$ &   O8 Iab & $70.0$ & $34.0$ & $34000$ & 
   $12.0\pe$ & $10.0\tr$ & & $18.0$ & $1500\pm200$ & $1370$ & $1700$ \\
&$152424$ &  O9.7 Ia & $52.0$ & $33.4$ & $30500$ & 
   $3.8\tr$ & $5.5\tr$ & & $6.5$ & $1500\pm100$ & $1350$ & $1540$ \\
63 Oph &$162978$ &  O8 III & $40.0$ & $16.0$ & $37100$ & 
   $0.83\je$ & & & $2.5$ & $2200\pm200$ & $2090$ & $2160$ \\
&$163758$ &  O8 III & $50.0$ & $20.1$ & $37200$ & 
   & & $5.0\os$ & $5.7$ & $2100\pm200$ & $1890$ & $2100$ \\
HR 7589 &$188209$ & O9.5 Ib & $43.0$ & $27.6$ & $30900$ & 
   $3.4\dv$ & & & $2.4$ & $1600\pm100$ & $1450$ & $1470$ \\
&$190864$ & O7 III & $42.0$ & $14.0$ & $39200$ & 
   $1.5\ct$ & & & $2.8$ & $2300\pm200$ & $2340$ & $2370$ \\
&$210809$ &  O9 Ib & $38.0$ & $21.4$ & $32000$ & 
   $4.0\ct$ & & & $1.0$ & $2000\pm200$ & $1700$ & $2200$ \\
$\lambda$  Cep&$210839$ &  O6 Iab & $51.0$ & $19.6$ & $38200$ & 
   $5.3\ct$ & $2.1\tr$ & $1.5\os$ & $4.7$ & $2100\pm200$ & $1930$ & $2150$\\
&$218915$ & O9.5 Iab & $43.0$ & $27.6$ & $30900$ & 
   $1.1\je$ & & & $2.4$ & $1800\pm100$ & $1450$ & $1470$ \\
\hline
\end{tabular}
\end{center}
\end{table*}

Because in
future we intend to calculate mainly wind models of B stars, to test the
reliability of our assumptions we calculated NLTE wind models of stars
with spectral type relatively close to B, i.e. later O stars.
For this purpose we selected O stars with a spectral type O6 or later
from the set of Lamers et al. (\cite{lsl}, hereafter LSL).
We included only stars for which reliabile determination of their
mass-loss rate exists in the literature.
Stellar parameters (masses, radii and effective temperatures) taken from
LSL are listed in Table\,\ref{ohvezpar}.

\subsection{Terminal velocities}

\begin{figure}
\resizebox{\hsize}{!}{\includegraphics{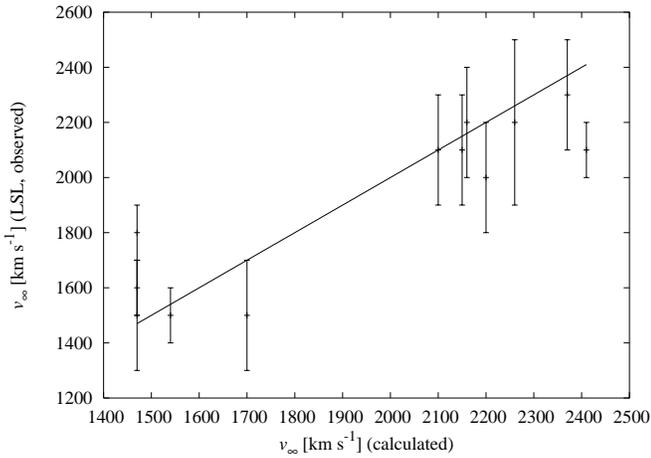}}
\caption[]{Comparison of calculated and observed (taken from LSL, BG)
terminal velocities.}
\label{vnek}
\end{figure}

\begin{figure}
\resizebox{\hsize}{!}{\includegraphics{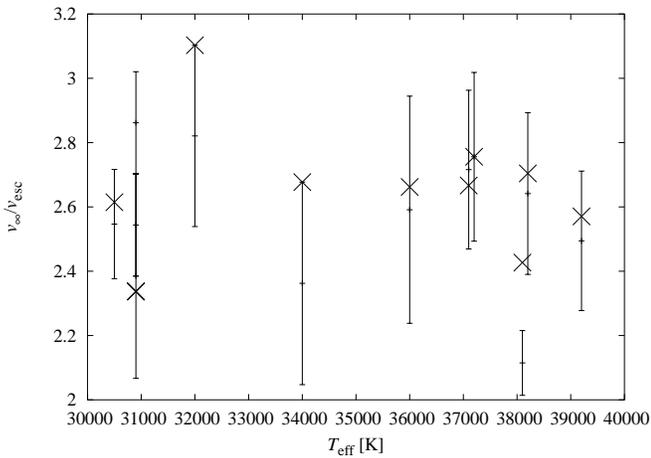}}
\caption[]{Comparison of calculated (crosses) and observed (taken from LSL, BG,
errorbars) ratio of the terminal velocity to the escape velocity
$v_\infty/v_\mathrm{esc}$.}
\label{vnekunik}
\end{figure}

Wind terminal velocities of galactic
O stars can be measured with a typical
uncertainty of $100\,\mathrm{km}\,\mathrm{s}^{-1}$.
Thus, in principle, they may provide 
a
relatively strict test for any wind
model. 
Unfortunately, because terminal velocities depend 
%
on surface
escape velocities (see Castor et al. \cite{cak}) the precise
prediction
of wind terminal velocity relies on the precise knowledge
of stellar mass and radius.
Consequently, the long-standing problem of ``mass discrepancy'' in
O~stars (cf.
Herrero et al. \cite{hekuku},
Lanz et al. \cite{hmotak}) complicates a straightforward
comparison of calculated and observed terminal velocities.
This is especially the case of stars \object{HD\,30614},
\object{HD\,188209}, and \object{HD\,218915},
for which we used the same stellar parameters, but which have different
observed terminal velocities.

The
comparison of observed (taken from LSL and Bianchi \& Garcia
\cite{bg}, hereafter BG) and calculated terminal velocities for model
stars shown in 
Fig.~\ref{vnek} implies that our
NLTE wind models are able to predict relatively correct values of wind
terminal velocities.
The fit is 
%
improved compared to our previous calculations
presented in KKII (see Table~\ref{ohvezpar}). Note
that our calculations consistently solved the problem of high
theoretical terminal velocities obtained by LSL using the so called
``cooking formula'' of Kudritzki et al. (\cite{varic}),
see also Table\,2 and Fig.\,12 in KKII.
However, since our models include several simplifying assumptions, 
it is necessary to test this conclusion against more advanced models.

It
remains unclear
%
why LSL obtained too high values
of terminal velocities when theoretical calculations (KKII) with
{\em the same} force multipliers yielded
terminal velocities consistent with observations.
Note that 
%
Pauldrach et al.
(\cite{pahole}) 
also
predicted correct terminal velocities, however for
a different set of stars.

The calculated terminal velocities are, in fact, not too 
%
different
from those
calculated by KKII, although the terminal velocities presented by KKII
were obtained simply using 
the
force multipliers of Abbott (\cite{abpar}).
This
relatively surprising conclusion
can be attributed to the fact that
the $\alpha$ parameter calculated properly using line statistics
(Puls et al. \cite{pusle}) is very close to the values calculated by
Abbott (\cite{abpar}).
And because the terminal velocities 
depend on the $\alpha$
value, terminal velocities calculated in this paper
and those
using Abbott's multipliers (KKII) do not differ significantly.
However, properly calculated $k$ parameters (Puls et al.
\cite{pusle}) are significantly lower than
those
calculated by Abbott (\cite{abpar}).
Indeed, our mass-loss rates (which are, in fact, influenced by 
the
$k$
parameter) are systematically lower than
those
of KKII and are close to the observed values (see also the next section
for the discussion of mass-loss rates).

The stellar wind theory predicts
a
strong correlation between wind terminal velocity and stellar escape
velocity. Thus, we have also compared
the ratio of the observed and calculated
terminal velocity to the escape velocity $v_\infty/v_\mathrm{esc}$
(see Fig. \ref{vnekunik}, $v_\mathrm{esc}$ is calculated using stellar
parameters from Table \ref{ohvezpar}).
Our derived mean
ratio $v_\infty/v_\mathrm{esc} \sim 2.6$
is in a good agreement with 
the
mean observed
$v_\infty/v_\mathrm{esc}$ ratio $2.5$ 
derived
%
by LSL for stars with
an effective temperature higher than $25\,000$\,K.
%

\subsection{Mass loss rates}

\begin{figure}
\resizebox{\hsize}{!}{\input{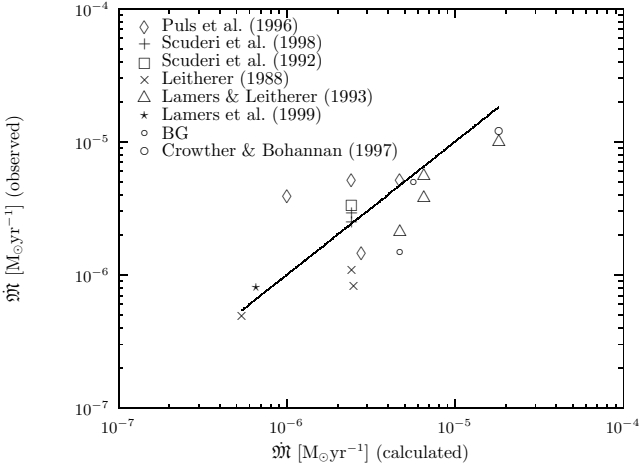}}
\caption[]{Comparison of calculated and observed mass-loss rates.}
\label{dmdt}
\end{figure}

\begin{figure}
\resizebox{\hsize}{!}{\input{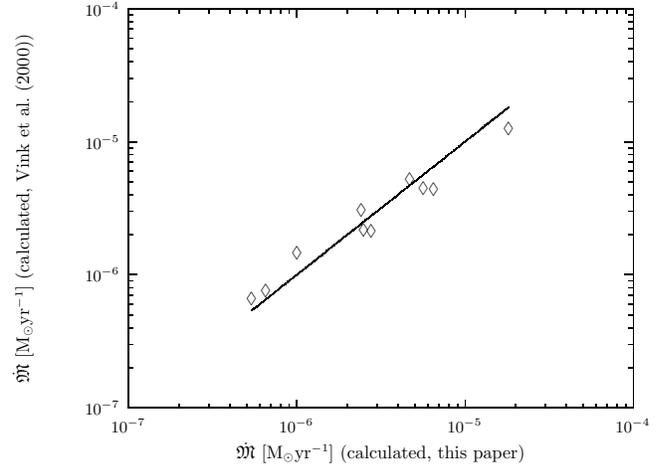}}
\caption[]{Comparison of theoretical mass-loss rates.}
\label{dmdtvkl}
\end{figure}

The credibility of observed mass-loss rates in most 
%
stars is much
worse than 
%
their wind terminal velocities.
Although in general the wind models are able to predict correct
mass-loss rates (cf. Vink et al. \cite{vikola}), there is a large
scatter for individual stars.
This situation is demonstrated in Fig.~\ref{dmdt} where we compare
calculated and observed mass-loss rates.
We compiled observed mass-loss rates based on the H$\alpha$ measurement
(Leitherer \cite{lei}, Scuderi et al. \cite{scupa}, Puls et al.
\cite{pulmoc}, Crowther \& Bohannan \cite{croboh}, Scuderi et al.
\cite{scupamoc},  Lamers et al.  \cite{lamoc}) and on the radio emission
(Lamers \& Leitherer \cite{lamlei}, Scuderi et al. \cite{scupamoc})
together with mass-loss determinations obtained using UV lines by BG.
Generally, UV 
line-diagnostic
is believed to be badly affected by the
unknown wind ionization fractions (Lamers et al. \cite{lamoc}).
This is especially the case of the O star winds for which unsaturated
wind lines originate only from trace elements.
However, precise modelling can overcome these difficulties (BG).

We conclude that there is generally a good agreement between predicted
and observed mass-loss rates.
However, significant differencies exist for individual stars.
Similarly, comparison of mass-loss rates calculated in this paper and by
Vink et al. (\cite{vikola}) shows relative good agreement between these
predicted mass-loss rates.
This possibly
supports the conclusion of Puls (\cite{puroz}) that multiple
scattering (included into Vink et al. (\cite{vikola}) models) does not
significantly influence mass-loss rates of weaker stellar winds.
%

An important caveat has to be mentioned.
The observed mass-loss rates may be influenced by wind
clumping 
(e.g.
Crowther et al. (\cite{cropul}), Massa et al. (\cite{mapul}),
%
Bouret et al. 
(\cite{boula}))
and thus, the observed
mass-loss rates shall be lower in such a case (see also Kramer et al.
\cite{kraco} for the discussion of effect of clumping on the theoretical
profiles of X-ray lines).
However, in such a case clumping would influence also
the theoretical mass-loss rates (cf. Gr\"afener \cite{gratu}).
%

\subsection{Wind momentum luminosity relationship}

\begin{figure}
\resizebox{\hsize}{!}{
\setlength{\unitlength}{0.240900pt}
\ifx\plotpoint\undefined\newsavebox{\plotpoint}\fi
\sbox{\plotpoint}{\rule[-0.200pt]{0.400pt}{0.400pt}}%
\begin{picture}(1500,1080)(0,0)
\font\gnuplot=cmr10 at 10pt
\gnuplot
\sbox{\plotpoint}{\rule[-0.200pt]{0.400pt}{0.400pt}}%
\put(201.0,123.0){\rule[-0.200pt]{4.818pt}{0.400pt}}
\put(181,123){\makebox(0,0)[r]{ 27}}
\put(1419.0,123.0){\rule[-0.200pt]{4.818pt}{0.400pt}}
\put(201.0,254.0){\rule[-0.200pt]{4.818pt}{0.400pt}}
\put(181,254){\makebox(0,0)[r]{ 27.5}}
\put(1419.0,254.0){\rule[-0.200pt]{4.818pt}{0.400pt}}
\put(201.0,385.0){\rule[-0.200pt]{4.818pt}{0.400pt}}
\put(181,385){\makebox(0,0)[r]{ 28}}
\put(1419.0,385.0){\rule[-0.200pt]{4.818pt}{0.400pt}}
\put(201.0,516.0){\rule[-0.200pt]{4.818pt}{0.400pt}}
\put(181,516){\makebox(0,0)[r]{ 28.5}}
\put(1419.0,516.0){\rule[-0.200pt]{4.818pt}{0.400pt}}
\put(201.0,647.0){\rule[-0.200pt]{4.818pt}{0.400pt}}
\put(181,647){\makebox(0,0)[r]{ 29}}
\put(1419.0,647.0){\rule[-0.200pt]{4.818pt}{0.400pt}}
\put(201.0,778.0){\rule[-0.200pt]{4.818pt}{0.400pt}}
\put(181,778){\makebox(0,0)[r]{ 29.5}}
\put(1419.0,778.0){\rule[-0.200pt]{4.818pt}{0.400pt}}
\put(201.0,909.0){\rule[-0.200pt]{4.818pt}{0.400pt}}
\put(181,909){\makebox(0,0)[r]{ 30}}
\put(1419.0,909.0){\rule[-0.200pt]{4.818pt}{0.400pt}}
\put(201.0,1040.0){\rule[-0.200pt]{4.818pt}{0.400pt}}
\put(181,1040){\makebox(0,0)[r]{ 30.5}}
\put(1419.0,1040.0){\rule[-0.200pt]{4.818pt}{0.400pt}}
\put(201.0,123.0){\rule[-0.200pt]{0.400pt}{4.818pt}}
\put(201,82){\makebox(0,0){ 4.6}}
\put(201.0,1020.0){\rule[-0.200pt]{0.400pt}{4.818pt}}
\put(339.0,123.0){\rule[-0.200pt]{0.400pt}{4.818pt}}
\put(339,82){\makebox(0,0){ 4.8}}
\put(339.0,1020.0){\rule[-0.200pt]{0.400pt}{4.818pt}}
\put(476.0,123.0){\rule[-0.200pt]{0.400pt}{4.818pt}}
\put(476,82){\makebox(0,0){ 5}}
\put(476.0,1020.0){\rule[-0.200pt]{0.400pt}{4.818pt}}
\put(614.0,123.0){\rule[-0.200pt]{0.400pt}{4.818pt}}
\put(614,82){\makebox(0,0){ 5.2}}
\put(614.0,1020.0){\rule[-0.200pt]{0.400pt}{4.818pt}}
\put(751.0,123.0){\rule[-0.200pt]{0.400pt}{4.818pt}}
\put(751,82){\makebox(0,0){ 5.4}}
\put(751.0,1020.0){\rule[-0.200pt]{0.400pt}{4.818pt}}
\put(889.0,123.0){\rule[-0.200pt]{0.400pt}{4.818pt}}
\put(889,82){\makebox(0,0){ 5.6}}
\put(889.0,1020.0){\rule[-0.200pt]{0.400pt}{4.818pt}}
\put(1026.0,123.0){\rule[-0.200pt]{0.400pt}{4.818pt}}
\put(1026,82){\makebox(0,0){ 5.8}}
\put(1026.0,1020.0){\rule[-0.200pt]{0.400pt}{4.818pt}}
\put(1164.0,123.0){\rule[-0.200pt]{0.400pt}{4.818pt}}
\put(1164,82){\makebox(0,0){ 6}}
\put(1164.0,1020.0){\rule[-0.200pt]{0.400pt}{4.818pt}}
\put(1301.0,123.0){\rule[-0.200pt]{0.400pt}{4.818pt}}
\put(1301,82){\makebox(0,0){ 6.2}}
\put(1301.0,1020.0){\rule[-0.200pt]{0.400pt}{4.818pt}}
\put(1439.0,123.0){\rule[-0.200pt]{0.400pt}{4.818pt}}
\put(1439,82){\makebox(0,0){ 6.4}}
\put(1439.0,1020.0){\rule[-0.200pt]{0.400pt}{4.818pt}}
\put(201.0,123.0){\rule[-0.200pt]{298.234pt}{0.400pt}}
\put(1439.0,123.0){\rule[-0.200pt]{0.400pt}{220.905pt}}
\put(201.0,1040.0){\rule[-0.200pt]{298.234pt}{0.400pt}}
\put(40,581){\makebox(0,0){\rotatebox{90}{$\log[\dmdt v_\infty (R_*/R_\odot)^{1/2}]$ (CGS units)}}}
\put(820,21){\makebox(0,0){$\log (L/L_\odot)$}}
\put(201.0,123.0){\rule[-0.200pt]{0.400pt}{220.905pt}}
\put(1102.0,829.0){\rule[-0.200pt]{0.400pt}{18.308pt}}
\put(1092.0,829.0){\rule[-0.200pt]{4.818pt}{0.400pt}}
\put(1092.0,905.0){\rule[-0.200pt]{4.818pt}{0.400pt}}
\put(1109.0,820.0){\rule[-0.200pt]{0.400pt}{18.308pt}}
\put(1099.0,820.0){\rule[-0.200pt]{4.818pt}{0.400pt}}
\put(1099.0,896.0){\rule[-0.200pt]{4.818pt}{0.400pt}}
\put(930.0,638.0){\rule[-0.200pt]{0.400pt}{18.308pt}}
\put(920.0,638.0){\rule[-0.200pt]{4.818pt}{0.400pt}}
\put(920.0,714.0){\rule[-0.200pt]{4.818pt}{0.400pt}}
\put(1295.0,884.0){\rule[-0.200pt]{0.400pt}{18.308pt}}
\put(1285.0,884.0){\rule[-0.200pt]{4.818pt}{0.400pt}}
\put(1285.0,960.0){\rule[-0.200pt]{4.818pt}{0.400pt}}
\put(758.0,539.0){\rule[-0.200pt]{0.400pt}{15.177pt}}
\put(748.0,539.0){\rule[-0.200pt]{4.818pt}{0.400pt}}
\put(748.0,602.0){\rule[-0.200pt]{4.818pt}{0.400pt}}
\put(1006.0,678.0){\rule[-0.200pt]{0.400pt}{16.381pt}}
\put(996.0,678.0){\rule[-0.200pt]{4.818pt}{0.400pt}}
\put(996.0,746.0){\rule[-0.200pt]{4.818pt}{0.400pt}}
\put(373.0,158.0){\rule[-0.200pt]{0.400pt}{16.381pt}}
\put(363.0,158.0){\rule[-0.200pt]{4.818pt}{0.400pt}}
\put(363.0,226.0){\rule[-0.200pt]{4.818pt}{0.400pt}}
\put(1033.0,873.0){\rule[-0.200pt]{33.244pt}{0.400pt}}
\put(1033.0,863.0){\rule[-0.200pt]{0.400pt}{4.818pt}}
\put(1171.0,863.0){\rule[-0.200pt]{0.400pt}{4.818pt}}
\put(1033.0,864.0){\rule[-0.200pt]{36.617pt}{0.400pt}}
\put(1033.0,854.0){\rule[-0.200pt]{0.400pt}{4.818pt}}
\put(1185.0,854.0){\rule[-0.200pt]{0.400pt}{4.818pt}}
\put(861.0,683.0){\rule[-0.200pt]{33.244pt}{0.400pt}}
\put(861.0,673.0){\rule[-0.200pt]{0.400pt}{4.818pt}}
\put(999.0,673.0){\rule[-0.200pt]{0.400pt}{4.818pt}}
\put(1219.0,929.0){\rule[-0.200pt]{36.376pt}{0.400pt}}
\put(1219.0,919.0){\rule[-0.200pt]{0.400pt}{4.818pt}}
\put(1370.0,919.0){\rule[-0.200pt]{0.400pt}{4.818pt}}
\put(696.0,570.0){\rule[-0.200pt]{29.872pt}{0.400pt}}
\put(696.0,560.0){\rule[-0.200pt]{0.400pt}{4.818pt}}
\put(820.0,560.0){\rule[-0.200pt]{0.400pt}{4.818pt}}
\put(937.0,712.0){\rule[-0.200pt]{33.003pt}{0.400pt}}
\put(937.0,702.0){\rule[-0.200pt]{0.400pt}{4.818pt}}
\put(1074.0,702.0){\rule[-0.200pt]{0.400pt}{4.818pt}}
\put(277.0,192.0){\rule[-0.200pt]{46.253pt}{0.400pt}}
\put(277.0,182.0){\rule[-0.200pt]{0.400pt}{4.818pt}}
\put(1102,873){\rule{1pt}{1pt}}
\put(1109,864){\rule{1pt}{1pt}}
\put(930,683){\rule{1pt}{1pt}}
\put(1295,929){\rule{1pt}{1pt}}
\put(758,570){\rule{1pt}{1pt}}
\put(1006,712){\rule{1pt}{1pt}}
\put(373,192){\rule{1pt}{1pt}}
\put(469.0,182.0){\rule[-0.200pt]{0.400pt}{4.818pt}}
\put(1022,666){\makebox(0,0){$+$}}
\put(660,493){\makebox(0,0){$+$}}
\put(722,521){\makebox(0,0){$+$}}
\put(1261,924){\makebox(0,0){$+$}}
\put(1121,795){\makebox(0,0){$+$}}
\put(915,682){\makebox(0,0){$+$}}
\put(1054,785){\makebox(0,0){$+$}}
\put(901,698){\makebox(0,0){$+$}}
\put(912,597){\makebox(0,0){$+$}}
\put(1071,765){\makebox(0,0){$+$}}
\put(277,207){\usebox{\plotpoint}}
\put(277.00,207.00){\usebox{\plotpoint}}
\put(294.24,218.54){\usebox{\plotpoint}}
\put(311.07,230.68){\usebox{\plotpoint}}
\put(328.27,242.29){\usebox{\plotpoint}}
\put(345.05,254.49){\usebox{\plotpoint}}
\put(362.18,266.20){\usebox{\plotpoint}}
\put(379.08,278.24){\usebox{\plotpoint}}
\put(396.25,289.89){\usebox{\plotpoint}}
\put(413.11,301.99){\usebox{\plotpoint}}
\put(430.32,313.57){\usebox{\plotpoint}}
\put(447.13,325.73){\usebox{\plotpoint}}
\put(463.92,337.94){\usebox{\plotpoint}}
\put(481.34,349.23){\usebox{\plotpoint}}
\put(498.30,361.19){\usebox{\plotpoint}}
\put(515.52,372.74){\usebox{\plotpoint}}
\put(532.37,384.87){\usebox{\plotpoint}}
\put(549.55,396.49){\usebox{\plotpoint}}
\put(566.34,408.70){\usebox{\plotpoint}}
\put(583.47,420.39){\usebox{\plotpoint}}
\put(600.36,432.45){\usebox{\plotpoint}}
\put(617.54,444.07){\usebox{\plotpoint}}
\put(634.39,456.19){\usebox{\plotpoint}}
\put(651.62,467.76){\usebox{\plotpoint}}
\put(668.42,479.94){\usebox{\plotpoint}}
\put(685.21,492.13){\usebox{\plotpoint}}
\put(702.44,503.69){\usebox{\plotpoint}}
\put(719.59,515.38){\usebox{\plotpoint}}
\put(736.81,526.95){\usebox{\plotpoint}}
\put(753.66,539.06){\usebox{\plotpoint}}
\put(770.83,550.70){\usebox{\plotpoint}}
\put(787.62,562.91){\usebox{\plotpoint}}
\put(804.77,574.58){\usebox{\plotpoint}}
\put(821.65,586.65){\usebox{\plotpoint}}
\put(838.84,598.26){\usebox{\plotpoint}}
\put(855.67,610.40){\usebox{\plotpoint}}
\put(872.91,621.94){\usebox{\plotpoint}}
\put(889.70,634.15){\usebox{\plotpoint}}
\put(906.51,646.32){\usebox{\plotpoint}}
\put(923.73,657.89){\usebox{\plotpoint}}
\put(940.58,670.00){\usebox{\plotpoint}}
\put(957.98,681.32){\usebox{\plotpoint}}
\put(974.96,693.24){\usebox{\plotpoint}}
\put(992.12,704.90){\usebox{\plotpoint}}
\put(1008.90,717.11){\usebox{\plotpoint}}
\put(1026.06,728.77){\usebox{\plotpoint}}
\put(1042.93,740.86){\usebox{\plotpoint}}
\put(1060.13,752.45){\usebox{\plotpoint}}
\put(1076.96,764.60){\usebox{\plotpoint}}
\put(1094.20,776.14){\usebox{\plotpoint}}
\put(1110.98,788.35){\usebox{\plotpoint}}
\put(1127.80,800.51){\usebox{\plotpoint}}
\put(1145.01,812.10){\usebox{\plotpoint}}
\put(1161.87,824.19){\usebox{\plotpoint}}
\put(1179.04,835.84){\usebox{\plotpoint}}
\put(1196.01,847.76){\usebox{\plotpoint}}
\put(1213.37,859.09){\usebox{\plotpoint}}
\put(1230.16,871.30){\usebox{\plotpoint}}
\put(1247.33,882.94){\usebox{\plotpoint}}
\put(1264.18,895.04){\usebox{\plotpoint}}
\put(1281.40,906.62){\usebox{\plotpoint}}
\put(1298.21,918.79){\usebox{\plotpoint}}
\put(1315.45,930.33){\usebox{\plotpoint}}
\put(1332.24,942.54){\usebox{\plotpoint}}
\put(1349.07,954.68){\usebox{\plotpoint}}
\put(1366.27,966.28){\usebox{\plotpoint}}
\put(1370,969){\usebox{\plotpoint}}
\end{picture}}
\caption[]{Comparison of calculated (crosses) and observed (errorbars)
 wind momentum (determined by Herrero et al. (\cite{hepun}) for Cyg OB2
association).
Linear regression of calculated data is also plotted.}
\label{wmlr}
\end{figure}
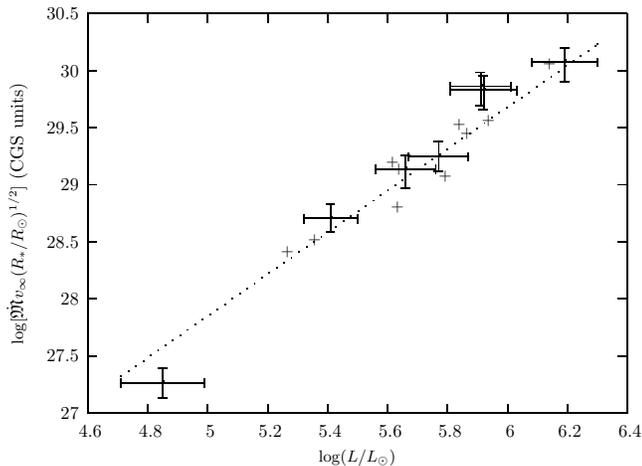

The wind momentum-luminosity relationship 
(see Kudritzki \& Puls \cite{kupul})
%
may provide a possibility to obtain precise measurements
of stellar and galaxy distances.
Thus, the exact calibration of this relationship is necessary.
Moreover, the wind momentum depends on the stellar mass only marginally,
and thus it provides a more reliable tool for the investigation of the
consistency between theoretical models and reality (Puls et al.
\cite{pulmoc}).

We have compared a theoretical modified wind momentum-luminosity
relationship $ \dot{\eu{M}} v_\infty \zav{R/R_{\sun}}^{1/2}$ 
obtained for
considered stars with our NLTE wind models with a relationship derived
from observations of supergiants
by Herrero et al. (\cite{hepun}) for the Cyg OB2
association (see Fig.~\ref{wmlr}).
Again, there is a relatively good agreement between calculated and
observed data.

\begin{table}[t]
\centering
\caption{Comparison of wind momentum-luminosity relationship from
different sources, both theoretical and observational
(see Eq. (\ref{rovmomlumvztah})). Parameters of the
sample denoted as "supergiants" were obtained by Repolust et al.
(\cite{rep}) combining their own results and results by Herrero et al.
(\cite{hepun}) for supergiants.
Similarly
values of the sample "giants and dwarfs" were obtained.
} 
\label{wmlrtab}
\begin{tabular}{lcc}
\hline
\hline
Sample & $\log D_0$ & $x$ \\
\hline
supergiants & $17.98 \pm 1.88$ & $ 2.00 \pm 0.32$\\
giants \& dwarfs & $18.70 \pm 1.29$ & $1.84 \pm 0.23$\\
Vink et al. (\cite{vikola}) & $18.68 \pm 0.26$ &$ 1.826 \pm 0.044$\\
this paper & $18.7\pm2.3$ & $1.83\pm0.40$\\
\hline
\end{tabular}
\end{table}

However, recent observational studies (Repolust et al. \cite{rep},
Markova et al. \cite{upice}) indicate that the situation is probably
more complicated. 
They concluded that there are two different wind momentum-luminosity
relationships for supergiants (or, probably more correctly, for stars
with the H$\alpha$ line
in emission), and for other stars (mainly for giants and main-sequence
stars).
However, theoretical studies (cf. Vink et al. \cite{vikola}) do not
predict such a difference.
Also results from our sample do not exhibit any significant difference
between wind momentum-luminosity relationships
for supergiant and for non-supergiant stars.
Repolust et al. (\cite{rep}) conclude that the observed difference may
be a consequence of clumping and that observed mass-loss rates of
high-density winds (mostly supergiant winds) may be artificially
enhanced by clumping.
To test this idea we calculated our predicted wind momentum-luminosity
relationship and compared it with both the
theoretical one (Vink et al. \cite{vikola}) and with the
observed wind momentum-luminosity relationship for
supergiants and non-supergiant stars (see Table~\ref{wmlrtab}).
Apparently, our predicted regression to the linear
wind momentum-luminosity 
relationship
\begin{equation}
\label{rovmomlumvztah}
\log\hzav{\dot{\eu{M}}v_\infty\zav{R/R_{\sun}}^{1/2}}=x\log(L/L_\odot)+\log D_0,
\quad \text{(CGS)}
\end{equation}
where $x$ and $D_0$ are fit parameters,
agrees well with the theoretical results of Vink et al. (\cite{vikola})
and with the observed relationship for non-supergiant stars.
However, the observed wind momentum-luminosity relationship for
supergiants is different. This confirms the finding
of Repolust et al. (\cite{rep}) that 
the
theoretical wind
momentum-luminosity relationship corresponds to the
observed one for non-supergiant stars.
Note, however, that most 
%
supergiants observed by Herrero et al.
(\cite{hepun}) represent 
exceptions
to this rule since most of them 
do not have an
H$\alpha$ line in emission.

\section{Discussion}
\label{disko}

The main improvement of the presented models is that the radiative force
is calculated without simplifying assumptions about the
line-distribution function (Puls et al. \cite{pusle}),
i.e., the parameters $k$, $\alpha$, and $\delta$ (Castor et al.
\cite{cak}, Abbott \cite{abpar}) are no longer required.
A correctly-calculated 
wind temperature structure influences occupation
numbers and thus, also the radiative force.
Moreover, these models are, in principle, able to consistently
calculate 
a
multicomponent wind structure, which is not necessary for the
case of O star winds, but 
that
will be invaluable for calculations of
low-density B star winds.

On the other hand, there are several assumptions involved in the
calculation of these models which can 
%
affect their validity.

First,
%
we assume 
a
stationary and spherically symmetric stellar
wind.
Both of these assumptions put severe limitations 
on
our models.
Variable phenomena, which are often observed in spectra of O stars,
cannot
be properly described with a stationary spherically symmetric
model of a wind. 
Spherical symmetry also prevents 
a
proper description of rotating winds.
On the other hand, a detailed numerical study of wind stability (cf.
Feldmeier at al. \cite{felpulpal}, Runacres \& Owocki \cite{runow}),
and of effects of stellar rotation on the stellar wind (cf. Petrenz \&
Puls \cite{ppjet}) showed that basic wind parameters (i.e. terminal
velocity and mass-loss rate) can be well reproduced by stationary and
spherically symmetric wind models (for rotational velocities lower than
the critical rotational velocity). However, wind clumping may affect
the values of theoretical mass-loss rates (Gr\"afener \cite{gratu}).
Finally, shocks caused by wind instabilities or by the presence of
strong magnetic fields (ud-Doula \& Owocki \cite{asta}) are sources of
strong X-ray and EUV emission.
This emission influences the ionization balance due to
the Auger ionization (Cassinelli \& Olson \cite{casol}) and direct
photoionization (Pauldrach et al. \cite{lepsipasam}).
On the other hand, MacFarlane et al. (\cite{macown}) showed, that for
O~stars the X-ray radiation influences the
ionization balance of trace elements only.
Thus, we conclude that at least wind
models of hotter star from our sample
are not significantly influenced by X-ray shock emission.

Another sources of possible errors are included in the approximate form
of model equations itself.
First, multiple scattering and line overlap may influence the mass-loss
rate, especially for high-density winds (see Abbott \& Lucy
\cite{abblu}, Puls \cite{puroz}, Vink et al. \cite{vikolabis}).
However, we plan to apply our code first to models of low-density winds,
for which the effect of multiple scattering is not important.

Second, radiative transfer in lines may be influenced by the continuum
effects. To overcome this limitation, we plan to include
continuum radiation 
in
the formulation of the radiative transfer
equation in 
the
Sobolev approximation
into our models after Hummer \& Rybicki (\cite{humrsrybou}).
For some lines, specially for the strong ones, the Sobolev approximation
may not be credible.
Thus, we plan to test the reliability of the
application of the Sobolev approximation for the calculation of the
radiative force by using a correct solution of the radiative
transfer equation in moving media (Kor\v c\'akov\'a \& Kub\'at
\cite{korku}).   This will
also allow as to account for the self-shadowing by photospheric lines
which is important for thin winds (Babel \cite{babelb}, together with
calculation of model atmospheres with inclusion of more elements).
Moreover, 
it
will enable a study of
instabilities connected with the
source-function gradients (Owocki \& Puls \cite{owpu}).
%

The effect of 
%
wind blanketing (Abbott \& Hummer \cite{wb}) was
neglected in our calculations to make the treatment of the wind easier.
However, it is clear that this effect causes backwarming of stellar
surface and may lead to different values of effective temperatures 
(e.g.
%
Herrero et al. \cite{hepun}).

The lambda-iterations employed for the solution of rate equations
together with radiative transfer in continuum may be source of another
potentially very serious problem.
However, the stellar wind is optically thin in most continuum
frequencies and the optically thick region occurs mostly relatively
near to the star.
On the other hand, the radiative transfer in lines in a Sobolev
approximation is solved consistently with rate equations.
Thus, due to these
reasons and due to a relatively good convergence of
our models we conclude that we are able to calculate consistent
occupation 
numbers
and wind structure even with this approximation.

The
processes of ionization and excitation can be very slow
compared to the typical flow-time.
Thus, 
the
ionization state of the wind can be "frozen" in a gas parcel and,
consequently, can be advected out.
Such a case 
%
in the stationary spherically symetric stellar
wind 
is
described by the rate equation in the form of
(Mihalas \cite{mihalas})
\begin{equation}
\frac{1}{r^2}\frac{\de\zav{r^2 n_i \io\vr}}{\de r} =
\sum_{j\neq i}n_j P_{ji}-n_i \sum_{j\neq i}P_{ij}=0,
\end{equation}
instead of Eq.\eqref{nlte}.
However, numerical tests showed that this phenomenon can be neglected
for most 
%
considered levels
(see also Lamers \& Morton \cite{lamo}).
Moreover, there would not be any radiatively driven wind if the
opposite is true since radiatively driven stellar wind is enabled by
high transition rates $P_{ij}$ (Gayley \cite{gayq}).
On the other hand, at first sight it is not clear whether this
assumption is also fulfilled in the case of strong wind-shocks.

Different 
approximations
of NLTE model atoms may influence
the
wind parameters obtained.
Besides the well-known errors in the atomic databases, inclusion of
superlevels may badly 
deteriorate
calculated models.
Whereas the effect of our simplified inclusion of superlevels is found
to be small, neglected collisional transitions between individual levels
forming the superlevel may affect wind temperature structure
(cf. Hillier \& Miller \cite{hilmi}).
Moreover, another subtle basic effect may influence the calculated
occupation numbers.
The calculated ionization fractions depend on the partition function
approximation.
Especially, in the relatively thin wind environment (where a large
number of higher excited levels have high occupation probabilities) 
%
different partition function approximations result in different final
ionization fractions.
Thus, it would be convenient to include more sophisticated methods for
the calculation of partition functions (cf. Hummer \& Mihalas
\cite{humi}) 
in
the model equations than those mentioned in 
Sect.~\ref{kapmodat}. 

\section{Conclusions}

We have presented new models of line-driven winds of hot stars.
The radiative force and the radiative heating term are consistently
calculated using statistical equilibrium equations.
The statistical equilibrium equations are solved for 15 most abundant
elements.
Obtained occupation numbers are used for the calculation of the
radiative force 
(in total
about 180\,000 lines are accounted for
in
the
calculation of the radiative force) and for the calculation of radiative
cooling and heating term.
All bound-bound collisional transitions and bound-free collisional and
radiative transitions explicitly included in the models are allowed to
contribute to the cooling and heating
terms.
Moreover, free-free transitions of H and He are also included.
 
We test the ability of our new models to predict correct wind parameters
of late O stars.
In particular, we compare calculated terminal velocities, mass-loss
rates and wind momenta with the observed ones.
We show that wind parameters calculated using our models are consistent
with observation.
Moreover, our new wind models have solved the problem introduced by LSL
of high theoretical wind terminal velocities. 
However, this conclusion 
has to
be tested again
against more advanced models, because our models involve several
simplifying assumptions. Moreover,
the relatively good correspondence between observed and
theoretical wind parameters can be 
reduced
by our insufficient
knowledge of stellar and wind parameters.

There is a good agreement between our calculated wind
momentum-luminosity relationship and that
relationship observed by Herrero et al. (\cite{hepun}), and Repolust et
al. (\cite{rep}) for non-supergiant stars.
However, contrary to the theoretical prediction of Vink et al. (\cite{vikola})
and ours, there is a difference between 
the
wind
momentum-luminosity relationship observed for non-supergiant stars and
for supergiants (or, more precisely, between stars with a different type
of H$\alpha$ profile). Repolust et al. (\cite{rep}) attribute
this difference to wind clumping.

On the other hand, many steps remains to be done.
Besides the overall improvements of our code we have to compare observed
and calculated wind parameters for B stars before we start to study
multicomponent models.

We
postpone detailed discussion of wind temperature structure
for 
a
separate paper.

\begin{acknowledgements}
The authors would like to thank Dr. J. Puls for his valuable comments on the
manuscript.
This research has made use of NASA's Astrophysics Data System and the
SIMBAD database, operated at CDS, Strasbourg, France.
This work was supported by a PPARC Rolling Grant and
by grants GA \v{C}R 205/01/0656 and
205/02/0445.
The Astronomical Institute Ond\v{r}ejov is supported by projects
K2043105 and Z1003909.
\end{acknowledgements}

\appendix
\section{Atomic data for NLTE calculations}
\label{atdata}

\begin{table*}[t]
\caption{Details of atomic models for NLTE calculation}
\label{atdatatab}
\centering
\begin{tabular}{lrrrc}
\hline\hline
Ion & Levels & Superlevels & Sublevels & Source \\
\hline
 \ion{H}{i}   &   9  &   0   &     &  TLUSTY \\
  \ion{He}{i}  &  5   &   9   &     &  TLUSTY \\
  \ion{He}{ii} &  14  &   0   &     &  TLUSTY \\
  \ion{C}{ii}  &  10  &   4   &     &  TLUSTY \\
  \ion{C}{iii} &  16  &   7   &     &  TLUSTY \\
  \ion{C}{iv}  &  21  &   4   &     &  TLUSTY \\
  \ion{N}{ii}  &  10  &   4   &     &  TLUSTY \\
  \ion{N}{iii} &  25  &   7   &     &  TLUSTY \\
  \ion{N}{iv}  &  15  &   8   &     &  TLUSTY \\
  \ion{N}{v}   &  10  &   3   &     &  TLUSTY \\
  \ion{O}{ii}  &  36  &  14   &     &  TLUSTY \\
  \ion{O}{iii} &  20  &   9   &     &  TLUSTY \\
  \ion{O}{iv}  &  31  &   8   &     &  TLUSTY \\
  \ion{O}{v}   &  12  &   2   &     &  TLUSTY \\
  \ion{O}{vi}  &  15  &   5   &     &  TLUSTY \\
  \ion{Ne}{ii} &  11  &   4   &     &  TLUSTY \\
  \ion{Ne}{iii}&  12  &   2   &     &  TLUSTY \\
 \ion{Ne}{iv} &  10 &    2   &      & TLUSTY \\
 \ion{Ne}{v}  &  12 &    5   &   34 & OP \\
 \ion{Na}{ii} &  9  &    4   &   37 & OP \\
 \ion{Na}{iii}&  12 &    2   &   21 & OP \\
 \ion{Na}{iv} &  13 &    5   &   40 & OP \\
 \ion{Na}{v}  &   8 &    8   &   51 & OP \\
 \ion{Mg}{iii}&   9 &    5   &   66 & OP \\
 \ion{Mg}{iv} &  12 &    2   &   21 & OP \\
 \ion{Mg}{v}  &   9 &    4   &   24 & OP \\
\hline
\end{tabular}
\hspace{1cm}
\begin{tabular}{lrrrc}
\hline\hline  
Ion & Levels & Superlevels & Sublevels & Source \\
\hline
\ion{Al}{ii}  &  12 &   4 &      & TLUSTY \\
\ion{Al}{iii} &   8 &   6 &   31 & OP \\
\ion{Al}{iv}  &   9 &   5 &   35 & OP \\
\ion{Al}{v}   &   5 &  11 &   74 & OP \\
\ion{Si}{ii}  &   9 &   3 &      & TLUSTY \\
\ion{Si}{iii} &  12 &   0 &      & TLUSTY \\
\ion{Si}{iv}  &   7 &   6 &      & TLUSTY \\
\ion{S}{ii}   &  14 &   0 &      & TLUSTY \\
\ion{S}{iii}  &   8 &   2 &      & TLUSTY \\
\ion{S}{iv}   &  14 &   4 &   22 & OP \\
\ion{S}{v}    &  12 &   2 &   18 & OP \\
\ion{Ar}{iii} &  13 &  12 &   61 & OP \\
\ion{Ar}{iv}  &  12 &   7 &   33 & OP \\
\ion{Ar}{v}   &  10 &   6 &   22 & OP \\
\ion{Ca}{ii}  &  11 &   5 &   42 & OP \\
\ion{Ca}{iii} &   8 &   6 &   68 & OP \\
\ion{Ca}{iv}  &  10 &  10 &  206 & OP \\
\ion{Ca}{v}   &   9 &  13 &  409 & OP \\
\ion{Fe}{iii} &   1 &  28 &  805 & IP \& OP \\
\ion{Fe}{iv}  &   1 &  31 & 1728 & IP \\
\ion{Fe}{v}   &   1 &  29 & 1818 & IP \\
\ion{Fe}{vi}  &   8 &  19 &  710 & IP \\
\ion{Ni}{iii} &   0 &  36 &      & TLUSTY \\
\ion{Ni}{iv}  &   0 &  38 &      & TLUSTY \\
\ion{Ni}{v}   &   0 &  48 &      & TLUSTY \\
\ion{Ni}{vi}  &   0 &   1 &      & TLUSTY \\
\hline
\end{tabular}
\end{table*}

The source of all data used for the calculation of model atoms is given
in Table \ref{atdatatab}.
Most important models are taken from the TLUSTY web page
{\tt http://tlusty.gsfc.nasa.gov}.

These atomic models were complemented by data constructed for some light
elements using the Opacity Project (hereafter OP; Seaton \cite{top},
Luo \& Pradhan \cite{top1}, Seaton et al. \cite{topt}, Butler et al.
\cite{bumez}, Nahar \& Pradhan \cite{napra}) database
{\tt http://vizier.u-strasbg.fr/OP.html}.
Energy levels, photoionization cross sections and oscillator strengths
were taken from OP data.
To avoid complicated calculation of numerous resonances (which shall be,
in fact, accounted for using solution of radiative transfer equation in
moving medium), we applied linear regression of photoionization data.
For the collisional ionization we applied the so-called Seaton's formula
(Mihalas \cite{mihalas}, Eq. (5.78) therein) with the photoionization
cross section at the ionization edge taken from OP data.
Finally, for the collisional excitation we used the Van Regemorter
formula (Mihalas \cite{mihalas}, Eq. (5.75) therein, Van Regemorter
\cite{vana}).

Another important atom is iron.
For \ion{Fe}{iii} we used energy levels and oscillator strengths from
Nahar \& Pradhan (\cite{zel2}), photoionization cross sections from OP
(Sawey \& Berrington \cite{savej}), collision rate coefficients for
low-lying levels from Zhang (\cite{zel1}), and the Van Regemorter
formula for levels not included in Zhang (\cite{zel1}).
Radiative rates for \ion{Fe}{iv} and \ion{Fe}{v} were calculated using
data from Bautista \& Pradhan (\cite{zel5}) or Bautista (\cite{zel6}),
and collision excitation rates are again calculated using either data
from Zhang \& Pradhan (\cite{zel4}) or the Van Regemorter formula.
Finally, radiative cross sections (both bound-bound and bound-free)
for \ion{Fe}{vi} are taken from OP data and collisional rate
coefficients are taken from Chen \& Pradhan (\cite{zel3}).

\section{Details of linearization of rates}
\label{nlteres}

Solution of NLTE rate equations is performed using Newton-Raphson
iterations (the so called linearization).
Linearization matrix follows from 
Eq.~\eqref{nlteskut}, where
derivatives of $P_{ij}$ corresponding to the line transitions are
evaluated using Eqs.~\eqref{zarex}, \eqref{zarde}, and \eqref{prenoscar}
-- \eqref{zdrojcar} as
\begin{subequations}
\begin{align}
\frac{\partial P_{ij}}{\partial N_i}=&B_{ij}\hzav{
(1-\beta)\frac{\partial S_{ij}}{\partial N_i} -
 \frac{\partial\beta}{\partial N_i}S_{ij}+ 
 \frac{\partial\beta_c}{\partial N_i} I_c},\\
\frac{\partial P_{ij}}{\partial N_j}=&B_{ij}\hzav{
(1-\beta)\frac{\partial S_{ij}}{\partial N_j} -
 \frac{\partial\beta}{\partial N_j}S_{ij}+ 
 \frac{\partial\beta_c}{\partial N_j} I_c},\\
\frac{\partial P_{ji}}{\partial N_i}=&B_{ji}\hzav{
(1-\beta)\frac{\partial S_{ij}}{\partial N_i} -
 \frac{\partial\beta}{\partial N_i}S_{ij}+ 
 \frac{\partial\beta_c}{\partial N_i} I_c},\\
\frac{\partial P_{ji}}{\partial N_j}=&B_{ji}\hzav{
(1-\beta)\frac{\partial S_{ij}}{\partial N_j} -
 \frac{\partial\beta}{\partial N_j}S_{ij}+ 
 \frac{\partial\beta_c}{\partial N_j} I_c},
\end{align}
\end{subequations}
where
\begin{subequations}
\begin{align}
\frac{\partial S_{ij}}{\partial N_i}&=
-\frac{S_{ij}B_{ij}}{N_i B_{ij}-N_j B_{ji}},\\
\frac{\partial S_{ij}}{\partial N_j}&=
\frac{S_{ij}}{N_j }+\frac{S_{ij}B_{ji}}{N_i B_{ij}-N_j B_{ji}},\\
\frac{\partial\beta}{\partial N_k}&=
\frac{1}{2}\int_{\mu_*}^{1}d\mu\hzav{\frac{e^{-\tau_\mu}}{\tau_\mu}-
                                    \frac{1-e^{-\tau_\mu}}{\tau_\mu^2}}
\frac{\partial\tau_\mu}{\partial N_k}, \quad k=i,j,\\
\frac{\partial\beta_c}{\partial N_k}&=
\frac{1}{2}\int_{-1}^{1}d\mu\hzav{\frac{e^{-\tau_\mu}}{\tau_\mu}-
                                    \frac{1-e^{-\tau_\mu}}{\tau_\mu^2}}
\frac{\partial\tau_\mu}{\partial N_k}, \quad k=i,j,\\
\frac{\partial\tau_\mu}{\partial N_i}&=\frac{\tilde\tau_\mu}{g_i},\\
\frac{\partial\tau_\mu}{\partial N_j}&=-\frac{\tilde\tau_\mu}{g_j},\\
\tilde\tau_\mu&=\frac{\pi e^2}{m_\mathrm{e}\nu_{ij}} z_{\rm atom}n_\mathrm{H}
g_if_{ij} \frac{r}{\io\vr\zav{1+\sigma\mu^2}}.
\end{align}
\end{subequations}
Note that the derivatives of collisional rates with respect to $N_i$ 
are
zero.
Finally, the derivatives of normalization condition \eqref{jednicka}
with respect of all relative populations $N_i$ are
\begin{equation}
\frac{\partial P_{ki}}{\partial N_i}=1,
\end{equation}
where $k$ is index of level with the highest 
occupation
number.

Similarly, for the calculation of derivatives of relative number
densities $N_i$ with respect to fluid variables $n_\mathrm{e}$,
$\rho_\mathrm{p}$, $T_\mathrm{e}$, $\sigma$ and $\io\vr$ in
Eq.~\eqref{nlteder} it is necessary to evaluate quantities
${\partial P_{ij}}/{\partial h}$. 

The derivatives of excitation and deexcitation
rates with respect to electron density
$n_e$ follow from 
Eqs.
\eqref{zarex}, \eqref{zarde}, and \eqref{srazky}, and
are relatively simple,
\begin{subequations}
\begin{align}
\frac{\partial R_{ij}}{\partial \el n}&=\frac{\partial R_{ji}}{\partial \el n}
=0,\\
\frac{\partial C_{ij}}{\partial \el n}&=\frac{C_{ij}}{\el n},\\
\frac{\partial C_{ji}}{\partial \el n}&=\frac{C_{ji}}{\el n},
\end{align}
\end{subequations}
whereas for the calculation of the derivatives of the ionization rates
(Eqs.~(\ref{zarion}) and (\ref{srazky}))
it is necessary to account for the
dependency of Saha-Boltzmann ratio on the electron density
\begin{subequations}
\begin{align}
\frac{\partial R_{ij}}{\partial \el n} &=0,\\
\frac{\partial R_{ji}}{\partial \el n} &=\frac{R_{ji}}{\el n},\\
\frac{\partial C_{ij}}{\partial \el n}&=\frac{C_{ij}}{\el n},\\
\frac{\partial C_{ji}}{\partial \el n}&=\frac{2C_{ji}}{\el n}.
\end{align}
\end{subequations}

The derivatives of the excitation rates with respect to electron
temperature are also relatively simple,
\begin{subequations}
\begin{align}
\frac{\partial R_{ij}}{\partial \el T}&=\frac{\partial R_{ji}}{\partial \el T}
=0,\\
\label{dercup}
\frac{\partial C_{ij}}{\partial \el T}&=n_\mathrm{e}
  \frac{\partial \Omega_{ij} (T_\mathrm{e})}{\partial \el T},\\
\label{dercdown}
\frac{\partial C_{ji}}{\partial \el T}&=n_\mathrm{e}
\frac{\partial}{\partial \el T} \zav{\frac{N_i}{N_j}}^*
\Omega_{ij} (T_\mathrm{e})+n_\mathrm{e} \zav{\frac{N_i}{N_j}}^* 
\frac{\partial \Omega_{ij} (T_\mathrm{e})}{\partial \el T},
\end{align}
\end{subequations}
however, for the evaluation of derivatives of ionization rates it is
necessary to perform an integration
\begin{subequations}
\begin{align}
\frac{\partial R_{ij}}{\partial \el T}&=0,\\
\frac{\partial R_{ji}}{\partial \el T}&=
4\pi \frac{\partial}{\partial \el T} \zav{\frac{N_i}{N_k}}^* 
\int_{\nu_i}^{\infty}
\frac{\alpha_{i,\nu}}{h\nu}\hzav{\frac{2h\nu^3}{c^2}+J_\nu}
e^{-\frac{h\nu}{kT_\mathrm{e}}}
\,\de\nu \nonumber\\&+
4\pi \zav{\frac{N_i}{N_k}}^*\int_{\nu_i}^{\infty}
\alpha_{i,\nu}\hzav{\frac{2h\nu^3}{c^2}+J_\nu}e^{-h\nu/kT_\mathrm{e}}
\frac{\de\nu}{kT_\mathrm{e}^2}.
\end{align}
\end{subequations}
Derivatives of collisional rates are given by Eqs.~\eqref{dercup} and
\eqref{dercdown}.

Using the dependence of the radiative transfer equation in Sobolev
approximation on the velocity gradient $\sigma$ and on $\io\vr$ and
$\pr\rho$ via the fraction $y=\pr\rho/\io\vr$ we calculate derivatives
of excitation rates with respect to these variables, namely
\begin{subequations}
\begin{align}
\frac{\partial R_{ij}}{\partial \sigma}&=B_{ij}\hzav{
 \frac{\partial\beta_c}{\partial \sigma} I_c-
\frac{\partial\beta}{\partial \sigma}S_{ij}},\\
\frac{\partial R_{ji}}{\partial \sigma}&=B_{ji}\hzav{
\frac{\partial\beta_c}{\partial \sigma} I_c-
\frac{\partial\beta}{\partial \sigma}S_{ij}},\\
\frac{\partial R_{ij}}{\partial \pr\rho}&=\frac{1}{\io\vr}\frac{\partial
R_{ij}}{\partial y},\\
\frac{\partial R_{ji}}{\partial \pr\rho}&=\frac{1}{\io\vr}\frac{\partial
R_{ji}}{\partial y},\\
\frac{\partial R_{ij}}{\partial \io\vr}&=-\frac{y}{\io\vr}\frac{\partial
R_{ij}}{\partial y},\\
\frac{\partial R_{ji}}{\partial \io\vr}&=-\frac{y}{\io\vr}\frac{\partial
R_{ji}}{\partial y},
\end{align}
\end{subequations}
where 
\begin{subequations}
\begin{align}
\frac{\partial\beta}{\partial \sigma}&=
\frac{1}{2}\int_{\mu_*}^{1}d\mu\frac{\mu^2}{1+\sigma\mu^2}
   \hzav{\frac{1-e^{-\tau_\mu}}{\tau_\mu}-e^{-\tau_\mu}},\\
\frac{\partial\beta_c}{\partial \sigma}&=
\frac{1}{2}\int_{-1}^{1}d\mu\frac{\mu^2}{1+\sigma\mu^2}
   \hzav{\frac{1-e^{-\tau_\mu}}{\tau_\mu}-e^{-\tau_\mu}},\\
\frac{\partial R_{ij}}{\partial y}&=B_{ij}\hzav{
 \frac{\partial\beta_c}{\partial y} I_c-
\frac{\partial\beta}{\partial y}S_{ij}},\\
\frac{\partial R_{ji}}{\partial y}&=B_{ji}\hzav{
\frac{\partial\beta_c}{\partial y} I_c-
\frac{\partial\beta}{\partial y}S_{ij}},\\
\frac{\partial\beta}{\partial y}&=
\frac{1}{2}\int_{\mu_*}^{1}d\mu\hzav{e^{-\tau_\mu}-
                         \frac{1-e^{-\tau_\mu}}{\tau_\mu}}\frac{1}{y},\\
\frac{\partial\beta_c}{\partial y}&=
\frac{1}{2}\int_{-1}^{1}d\mu\hzav{e^{-\tau_\mu}-
                         \frac{1-e^{-\tau_\mu}}{\tau_\mu}}\frac{1}{y}.
\end{align}
\end{subequations}
Note that the
variable $\sigma$ is not an indepentend variable during linearization.
Derivatives with respect to $\sigma$ are transformed to derivatives of
ionic velocity using Eq. \eqref{sigma}.

\section{Details of linearization of raditive heating/cooling term}
\label{heacoo}

Radiative heating/cooling term in the case of thermal balance of
electrons is given by (Kub\'at et al. \cite{kpp})
\begin{equation}
\label{qrads}
Q^{\rad}=Q_{\mathrm{ff}}^{H}+Q_{\mathrm{bf}}^{H}+Q_{\mathrm{c}}^{H}-
         Q_{\mathrm{ff}}^{C}-Q_{\mathrm{bf}}^{C}-Q_{\mathrm{c}}^{C},
\end{equation}
where individual terms are
\begin{subequations}
\label{qradjed}
\begin{align}
Q_{\mathrm{ff}}^{H}\left[n_i\right]  = &  4\pi \el n \sum_{i=\ion{H}{ii}, \ion{He}{iii}}
  n_{i}\int_{0}^{\infty}
      \alpha_{\mathrm{ff},i}(\nu,\el T) J_{\nu}\de \nu,\\
Q_{\mathrm{ff}}^{C}\left[n_i\right]   = &  4\pi \el n \sum_{i=\ion{H}{ii}, \ion{He}{iii}}
   \int_{0}^{\infty}n_{i}
      \alpha_{\mathrm{ff}}(\nu,\el T) 
 \nonumber \times \\*& \times \zav{J_{\nu} + \frac{2h\nu^3}{c^2}}
      \er^{-{h\nu}/{k\el T}}\de \nu,\\
Q_{\mathrm{bf}}^{H}\left[n_i\right]   = & 4\pi \sum_{i} n_i 
   \int_{\nu_i}^{\infty}
\alpha_{i,\nu}
J_{\nu}\zav{1-\frac{\nu_{i}}{\nu}}\de \nu,\\
Q_{\mathrm{bf}}^{C}\left[n_{k_i}\right]   = & 4\pi \sum_{i} 
\int_{\nu_i}^{\infty} n_{k_i}\zav{\frac{N_i}{N_{k_i}}}^* \alpha_{i,\nu}
\zav{J_{\nu} + \frac{2h\nu^3}{c^2}}\nonumber \times \\&\times
       \er^{-h\nu/k\el T} 
\zav{1-\frac{\nu_{i}}{\nu}}\de \nu, \\
Q_{\mathrm{c}}^{H}\left[n_j\right]   = & \el n \sum_{ij} n_j \zav{\frac{N_i}{N_j}}^* \Omega_{ij}
(T_\mathrm{e}) h\nu_{ij}
,\\
Q_{\mathrm{c}}^{C}\left[n_i\right] = & \el n \sum_{ij} n_i \Omega_{ij} (T_\mathrm{e})
h\nu_{ij},
\end{align}
\end{subequations}
where $k_i$ denotes level into which is level $i$ ionized.
From these equations derivatives of heating/cooling term with respect to
$\io\vr$, $\pr\rho$, $\el\rho$, $\el T$ for the Newton-Raphson
iteration step of hydrodynamical structure can be calculated,
\begin{subequations}
\renewcommand{\theequation}{%
  \ifnum\value{equation}<27%
    \theparentequation\alph{equation}%
  \else\addtocounter{equation}{-26}%
       \theparentequation\Alph{equation}%
       \addtocounter{equation}{26}\fi}
\begin{align}
\frac{\partial Q_{\mathrm{ff}}^{H}}{\partial \el T}  = & 
    Q_{\mathrm{ff}}^{H}\left[\frac{\partial n_{i}}{\partial \el T}\right]-
    \frac{1}{2\el T} Q_{\mathrm{ff}}^{H}\left[n_i\right] ,\\
\frac{\partial Q_{\mathrm{ff}}^{H}}{\partial \el n}  = &
\frac{1}{\el n}Q_{\mathrm{ff}}^{H}\left[n_i\right]+
Q_{\mathrm{ff}}^{H}\left[\frac{\partial n_{i}}{\partial \el n}\right],\\
\frac{\partial Q_{\mathrm{ff}}^{H}}{\partial \pr \rho}  = &
Q_{\mathrm{ff}}^{H}\left[\frac{\partial n_{i}}{\partial \pr \rho}\right],\\
\frac{\partial Q_{\mathrm{ff}}^{H}}{\partial \io\vr}  = &
Q_{\mathrm{ff}}^{H}\left[\frac{\partial n_{i}}{\partial \io\vr}\right],\\
\frac{\partial Q_{\mathrm{ff}}^{H}}{\partial \sigma}  = &
Q_{\mathrm{ff}}^{H}\left[\frac{\partial n_{i}}{\partial\sigma}\right] ,\\
\frac{\partial Q_{\mathrm{ff}}^{C}}{\partial \el T}=
&Q_{\mathrm{ff}}^{C}\left[\frac{\partial n_{i}}{\partial \el T}\right]-
\frac{h}{k\el T}Q_{\mathrm{ff}}^{C}\left[\nu n_{i}\right]-
\frac{1}{2\el T}Q_{\mathrm{ff}}^{C}\left[n_{i}\right],\\
\frac{\partial Q_{\mathrm{ff}}^{C}}{\partial \el n}=&
\frac{1}{\el n}Q_{\mathrm{ff}}^{C}\left[n_i\right]+
Q_{\mathrm{ff}}^{C}\left[\frac{\partial n_{i}}{\partial \el n}\right],\\
\frac{\partial Q_{\mathrm{ff}}^{C}}{\partial\pr\rho}=&
Q_{\mathrm{ff}}^{C}\left[\frac{\partial n_{i}}{\partial \pr \rho}\right],\\
\frac{\partial Q_{\mathrm{ff}}^{C}}{\partial \io\vr}  = &
Q_{\mathrm{ff}}^{C}\left[\frac{\partial n_{i}}{\partial \io\vr}\right],\\
\frac{\partial Q_{\mathrm{ff}}^{C}}{\partial \sigma}  = &
Q_{\mathrm{ff}}^{C}\left[\frac{\partial n_{i}}{\partial\sigma}\right] ,\\
\frac{\partial Q_{\mathrm{bf}}^{H}}{\partial \el T}  = & 
    Q_{\mathrm{bf}}^{H}\left[\frac{\partial n_{i}}{\partial \el T}\right],\\
\frac{\partial Q_{\mathrm{bf}}^{H}}{\partial \el n}  = &
Q_{\mathrm{bf}}^{H}\left[\frac{\partial n_{i}}{\partial \el n}\right],\\
\frac{\partial Q_{\mathrm{bf}}^{H}}{\partial \pr \rho}  = &
Q_{\mathrm{bf}}^{H}\left[\frac{\partial n_{i}}{\partial \pr \rho}\right],\\
\frac{\partial Q_{\mathrm{bf}}^{H}}{\partial \io\vr}  = &
Q_{\mathrm{bf}}^{H}\left[\frac{\partial n_{i}}{\partial \io\vr}\right],\\
\frac{\partial Q_{\mathrm{bf}}^{H}}{\partial \sigma}  = &
Q_{\mathrm{bf}}^{H}\left[\frac{\partial n_{i}}{\partial\sigma}\right] ,\\
\nonumber
\frac{\partial Q_{\mathrm{bf}}^{C}}{\partial \el T}=
&Q_{\mathrm{bf}}^{C}\left[\frac{\partial n_{k_i}}{\partial\el T}\right]-
\frac{h}{k\el T}Q_{\mathrm{bf}}^{C}\left[\nu n_{k_i}\right]+\\*&+
Q_{\mathrm{bf}}^{C}\left[n_{k_i}\left\{\zav{\frac{N_i}{N_{k_i}}}^*\right\}^{-1}
\frac{\partial}{\partial \el T} \zav{\frac{N_i}{N_{k_i}}}^*\right],\\
\nonumber
\frac{\partial Q_{\mathrm{bf}}^{C}}{\partial \el n}=&
Q_{\mathrm{bf}}^{C}\left[\frac{\partial n_{k_i}}{\partial\el n}\right]+\\*&+
Q_{\mathrm{bf}}^{C}\left[n_{k_i}\left\{\zav{\frac{N_i}{N_{k_i}}}^*\right\}^{-1}
\frac{\partial}{\partial \el n} \zav{\frac{N_i}{N_{k_i}}}^*\right],\\
\frac{\partial Q_{\mathrm{bf}}^{C}}{\partial\pr\rho}=&
Q_{\mathrm{bf}}^{C}\left[\frac{\partial n_{k_i}}{\partial \pr \rho}\right],\\
\frac{\partial Q_{\mathrm{bf}}^{C}}{\partial \io\vr}  = &
Q_{\mathrm{bf}}^{C}\left[\frac{\partial n_{k_i}}{\partial \io\vr}\right],\\
\frac{\partial Q_{\mathrm{bf}}^{C}}{\partial \sigma}  = &
Q_{\mathrm{bf}}^{C}\left[\frac{\partial n_{k_i}}{\partial\sigma}\right] ,\\
\nonumber
\frac{\partial Q_{\mathrm{c}}^{H}}{\partial \el T}=&
Q_{\mathrm{c}}^{H}\left[\frac{\partial n_{j}}{\partial \el T}\right]+
\el n \sum_{ij} n_j\frac{\partial}{\partial \el T} \zav{\frac{N_i}{N_j}}^*
\Omega_{ij}(\el T)h\nu_{ij} +\\&+
\el n \sum_{ij} n_j\zav{\frac{N_i}{N_j}}^*
\frac{\partial\Omega_{ij}(\el T)}{\partial \el T}h\nu_{ij},\\
\nonumber
\frac{\partial Q_{\mathrm{c}}^{H}}{\partial \el n}=&
\frac{1}{\el n}Q_{\mathrm{c}}^{H}\left[n_{j}\right]+
Q_{\mathrm{c}}^{H}\left[\frac{\partial n_{j}}{\partial \el n}\right]+\\&+
\el n \sum_{ij} n_j\frac{\partial}{\partial \el n} \zav{\frac{N_i}{N_j}}^*
\Omega_{ij}(\el T)h\nu_{ij},\\
\frac{\partial Q_{\mathrm{c}}^{H}}{\partial \pr \rho}=&
Q_{\mathrm{c}}^{H}\left[\frac{\partial n_{j}}{\partial \pr \rho}\right],\\
\frac{\partial Q_{\mathrm{c}}^{H}}{\partial\io\vr}=&
Q_{\mathrm{c}}^{H}\left[\frac{\partial n_{j}}{\partial\io\vr}\right],\\
\frac{\partial Q_{\mathrm{c}}^{H}}{\partial\sigma}=&
Q_{\mathrm{c}}^{H}\left[\frac{\partial n_{j}}{\partial\sigma}\right],\\
\frac{\partial Q_{\mathrm{c}}^{C}}{\partial \el T}=&
Q_{\mathrm{c}}^{C}\left[\frac{\partial n_{i}}{\partial \el T}\right]+
\el n \sum_{ij} n_i
\frac{\partial\Omega_{ij}(\el T)}{\partial \el T}h\nu_{ij},\\
\frac{\partial Q_{\mathrm{c}}^{C}}{\partial \el n}=&
\frac{1}{\el n}Q_{\mathrm{c}}^{C}\left[n_{i}\right]+
Q_{\mathrm{c}}^{C}\left[\frac{\partial n_{i}}{\partial \el n}\right],\\
\frac{\partial Q_{\mathrm{c}}^{C}}{\partial \pr \rho}=&
Q_{\mathrm{c}}^{C}\left[\frac{\partial n_{i}}{\partial \pr \rho}\right],\\
\frac{\partial Q_{\mathrm{c}}^{C}}{\partial\io\vr}=&
Q_{\mathrm{c}}^{C}\left[\frac{\partial n_{i}}{\partial\io\vr}\right],\\
\frac{\partial Q_{\mathrm{c}}^{C}}{\partial\sigma}=&
Q_{\mathrm{c}}^{C}\left[\frac{\partial n_{i}}{\partial\sigma}\right] ,
\end{align}
\end{subequations}
where
\begin{subequations}
\begin{align}
\frac{\partial n_{i}}{\partial x}&=\frac{n_i}{N_i}
  \frac{\partial N_{i}}{\partial x},\qquad\text{where}\;x=\el T, \el n, \io\vr,
  \sigma,\\
\frac{\partial n_{i}}{\partial\pr\rho}&=\frac{n_i}{\pr\rho}+
  \frac{n_i}{N_i}\frac{\partial N_{i}}{\partial\pr\rho}.
\end{align}
\end{subequations}

\end{document}